\documentclass[a4paper,10pt,journal]{IEEEtran}
\makeatletter
\newcommand{\AddInputPath}[1]{%
  \ifx\input@path\@undefined
    \def\input@path{#1}
  \else
    \g@addto@macro{\input@path}{#1}
  \fi
}
\makeatother
\AddInputPath{{../}}

\usepackage{etex}

\usepackage[font=footnotesize,caption=false]{subfig} % preload with correct options...

\usepackage{relsize}

\usepackage{array}
\usepackage{booktabs,tabularx}
\usepackage{multirow}
\usepackage[binary-units=true,per-mode=symbol,detect-mode=true]{siunitx}
\usepackage{graphicx}

\usepackage[T1]{fontenc}
\usepackage{textcomp}
\usepackage[utf8]{inputenc}
\usepackage[final]{microtype}
\usepackage{icomma}
\usepackage{xspace}

\usepackage[tbtags]{amsmath}
\usepackage{amssymb,amsfonts,bm}
\usepackage{mathtools} 
\usepackage{dsfont}
\usepackage{mathrsfs}
\usepackage{accents}
\usepackage{empheq}
\usepackage{nccmath}
\usepackage{balance}

\usepackage{color}
\usepackage{calc}
\usepackage{tikz}
\usepackage{pgfplots,pgfplotstable}

\usepackage{pdftexcmds}
\makeatletter
\newcommand{\strequal}[2]{\pdf@strcmp{#1}{#2}==0}
\makeatother

\usepackage[capitalize]{cleveref}
\usepackage{refcount}

\usepackage[inline]{enumitem}
\usepackage{algorithm}
\usepackage{algpseudocode}
\makeatletter
\newcommand{\algmargin}{\the\ALG@thistlm}
\makeatother
\newlength{\whilewidth}
\algdef{SE}[parWHILE]{parWhile}{EndparWhile}[1]
  {\parbox[t]{\dimexpr\linewidth-\algmargin}{%
     \hangindent\whilewidth\strut\algorithmicwhile\ #1\ \algorithmicdo\strut}}{\algorithmicend\ \algorithmicwhile}%
\algnewcommand{\parState}[1]{\State%
  \parbox[t]{\dimexpr\linewidth-\algmargin}{\strut #1\strut}}

\makeatletter
\newcommand\fs@spaceruled{\def\@fs@cfont{\bfseries}\let\@fs@capt\floatc@ruled
  \def\@fs@pre{\vspace{.05in}\hrule height.8pt depth0pt \kern2pt}%
  \def\@fs@post{\kern2pt\hrule\relax}%
  \def\@fs@mid{\kern2pt\hrule\kern2pt}%
  \let\@fs@iftopcapt\iftrue}
\makeatother

\usepackage{glossaries}
\usepackage{ifthen}
\usepackage[noadjust]{cite}
\usepackage{multibib}

\usepackage{comment}
\usepackage{todonotes}
\let\legacytodo\todo
\newcommand{\ruggedtodo}[2][]{\tikzexternaldisable\legacytodo[#1]{#2}\tikzexternalenable}
\renewcommand{\todo}[1]{\ruggedtodo[inline]{#1}}

\makeglossaries

\newacronym{sca}{SCA}{successive convex approximation}
\newacronym{wmmse}{WMMSE}{weighted minimum mean squared error}
\newacronym{cdf}{CDF}{cumulative distribution function}
\newacronym{mulp}{MU-LP}{multi-user linear precoding}
\newacronym{scsic}{SC-SIC}{superposition coding and successive interference cancellation}
\newacronym{sc}{SC}{superposition coding}
\newacronym{cbs}{CBS}{current best solution}
\newacronym{cbv}{CBV}{current best value}
\newacronym{los}{LOS}{line-of-sight}
\newacronym{leo}{LEO}{low earth orbit}
\newacronym{iot}{IoT}{internet of things}
\newacronym{irs}{IRS}{intelligent reflecting surface}
\newacronym{socp}{SOCP}{second-order cone program}
\newacronym{soc}{SOC}{second-order cone}
\newacronym{dsl}{DSL}{digital subscriber line}
\newacronym{wsee}{WSEE}{weighted sum energy efficiency}
\newacronym{wsr}{WSR}{weighted sum rate}
\newacronym{mmwave}{mmWave}{millimeter wave}
\newacronym{dfg}{DFG}{Deutsche Forschungsgemeinschaft}
\newacronym{haec}{HAEC}{Highly Adaptive Energy-Efficient Computing}
\newacronym{hpc}{HPC}{High Performance Computing}
\newacronym{mac}{MAC}{multiple-access channel}
\newacronym{bc}{BC}{broadcast channel}
\newacronym{siso}{SISO}{single-input single-output}
\newacronym{simo}{SIMO}{single-input multiple-output}
\newacronym{miso}{MISO}{multiple-input single-output}
\newacronym{mimo}{MIMO}{multiple-input multiple-output}
\newacronym{af}{AF}{amplify-and-forward}
\newacronym{df}{DF}{decode-and-forward}
\newacronym{cf}{CF}{compress-and-forward}
\newacronym{mwrc}{MWRC}{multi-way relay channel}
\newacronym{dmmwrc}{DM-MWRC}{discrete memoryless multi-way relay channel}
\newacronym{pde}{PDE}{partial data exchange}
\newacronym{fde}{FDE}{full data exchange}
\newacronym{iid}{i.i.d.\@}{independent and identically distributed}
\newacronym{di}{DI} {difference of increasing}
\newacronym{dc}{DC}{difference of convex}
\newacronym{mm}{MM}{mixed monotonic}
\newacronym{mmp}{MMP}{mixed monotonic programming}
\newacronym{awgn}{AWGN}{additive white Gaussian noise}
\newacronym{wgn}{WGN}{white Gaussian noise}
\newacronym{awg}{AWG}{additive white Gaussian}
\newacronym{sic}{SIC}{successive interference cancellation}
\newacronym{snr}{SNR}{signal-to-noise ratio}
\newacronym{sinr}{SINR}{signal to interference plus noise ratio}
\newacronym{inr}{INR}{interference to noise ratio}
\newacronym{zf}{ZF}{zero-forcing}
\newacronym{mrt}{MRT}{maximum ratio transmission}
\newacronym{mmse}{MMSE}{minimum mean square error}
\newacronym{sud}{SUD}{single user decoding}
\newacronym{dof}{DoF}{degrees of freedom}
\newacronym{gdof}{GDoF}{generalized degrees of freedom}
\newacronym{nnc}{NNC}{noisy network coding}
\newacronym{dmn}{DMN}{discrete memoryless network}
\newacronym{csi}{CSI}{channel state information}
\newacronym{pmf}{pmf}{probability mass function}
\newacronym{dmic}{DM-IC}{discrete memoryless interference channel}
\newacronym{ic}{IC}{interference channel}
\newacronym{gic}{GIC}{Gaussian interference channel}
\newacronym{if}{IF}{interference}
\newacronym{ee}{EE}{energy efficiency}
\newacronym{gee}{GEE}{global energy efficiency}
\newacronym{tin}{TIN}{treating interference as noise}
\newacronym{snd}{SND}{simultaneous non-unique decoding}
\newacronym{sd}{SD}{simultaneous decoding}
\newacronym{hk}{HK}{Han-Kobayashi}
\newacronym{rs}{RS}{rate splitting}
\newacronym{rf}{RF}{radio frequency}
\newacronym{pa}{PA}{power amplifier}
\newacronym{lna}{LNA}{low noise amplifier}
\newacronym{lo}{LO}{local oscillator}
\newacronym{adc}{ADC}{analog-to-digital converter}
\newacronym{dac}{DAC}{digital-to-analog converter}
\newacronym{dsp}{DSP}{digital signal processing}
\newacronym{brd}{BRD}{best response dynamics}
\newacronym{br}{BR}{best response}
\newacronym{ne}{NE}{Nash equilibrium}
\newacronym{lhs}{LHS}{left-hand side}
\newacronym{rhs}{RHS}{right-hand side}
\newacronym{ran}{RAN}{radio access network}
\newacronym{qos}{QoS}{Quality of Service}
\newacronym{ngmn}{NGMN}{Next Generation Mobile Networks}
\newacronym{cap}{CAP}{Capacity Adaptation}
\newacronym{bwa}{BW}{Bandwidth Adaptation}
\newacronym{prb}{PRB}{physical resource block}
\newacronym{se}{SE}{spectral efficiency}
\newacronym{tp}{TP}{throughput}
\newacronym{bs}{BS}{base station}
\newacronym{ue}{UE}{user equipment}
\newacronym{mop}{MOP}{multi-objective optimization problem}
\newacronym{gda}{GDA}{generalized Dinkelbach's algorithm}
\newacronym{midcp}{MIDCP}{mixed integer disciplined convex programming}
\newacronym{lp}{LP}{linear program}
\newacronym{brb}{BRB}{branch reduce and bound}
\newacronym{bb}{BB}{branch and bound}
\newacronym{sit}{SIT}{successive incumbent transcending}
\newacronym{oma}{OMA}{orthogonal multiple access}
\newacronym{noma}{NOMA}{non-orthogonal multiple access}
\newacronym{rsma}{RSMA}{rate splitting multiple access}
\newacronym{sdma}{SDMA}{space division multiple access}
\newacronym{wlog}{w.l.o.g.\@}{without loss of generality}
\newacronym{lsc}{l.s.c.\@}{lower semi-continuous}
\newacronym{usc}{u.s.c.\@}{upper semi-continuous}
\newacronym{kkt}{KKT}{Karush-Kuhn-Tucker}
\newacronym{ptp}{PTP}{point-to-point}
\newacronym{phy}{PHY}{physical}
\newacronym{csit}{CSIT}{channel state information at the transmitter}
\newacronym{rbf}{RBF}{random beamforming}
\newacronym{mbf}{MBF}{matched beamforming}
\newacronym{svd}{SVD}{singular value decomposition}
\newacronym{zfbf}{ZFBF}{zero forcing beamforming}
\newacronym{rzf}{RZF}{reguralized-zero forcing beamforming}
\newacronym{bd}{BD}{block diagonalization}
\newacronym{admm}{ADMM}{alternating direction method of multipliers}
\newacronym{sdr}{SDR}{semidefinite relaxation}
\glsenableentrycount
\makeglossaries

\usetikzlibrary{positioning}
\usetikzlibrary{calc}
\usetikzlibrary{math}
\usetikzlibrary{fit}
\usetikzlibrary{intersections}
\usetikzlibrary{decorations.pathreplacing}
\usetikzlibrary{decorations.markings}
\usetikzlibrary{3d,angles}

\usetikzlibrary{external}
\tikzset{
	antenna/.pic={
		\draw[thick] (0,0) -- ++(120:2mm) -- ++(0:2mm) -- cycle -- (0,-1.5mm);
	}
}


\crefname{equation}{}{}
\crefrangeformat{equation}{(#3#1#4)--(#5#2#6)}
\crefmultiformat{equation}{(#2#1#3)}{ and~(#2#1#3)}{, (#2#1#3)}{, (#2#1#3)}
\crefrangemultiformat{equation}{#3(#1)#4--#5(#2)#6}{, #3(#1)#4--#5(#2)#6}{, #3(#1)#4--#5(#2)#6}{, #3(#1)#4--#5(#2)#6}
\DeclareMathOperator*\argmax{arg\,max}
\DeclareMathOperator*\argmin{arg\,min}

\undef\mod
\DeclareMathOperator\mod{mod}

\DeclareMathOperator\proj{proj}

\newcommand{\st}{\mathrm{s.\,t.}}

\let\vec\bm

\newcommand{\ubar}[1]{\underaccent{\bar}{#1}}

\allowdisplaybreaks[3]

\DeclareSIUnit \dBm {dBm}
\DeclareSIUnit \dBW {dBW}
\DeclareSIUnit \bpcu {bpcu}

\DeclareFontFamily{U}{mathx}{\hyphenchar\font45}
\DeclareFontShape{U}{mathx}{m}{n}{
      <5> <6> <7> <8> <9> <10>
      <10.95> <12> <14.4> <17.28> <20.74> <24.88>
      mathx10
      }{}
\DeclareSymbolFont{mathx}{U}{mathx}{m}{n}
\DeclareMathSymbol{\bigtimes}{1}{mathx}{"91}

\newtheorem{theorem}{Theorem}
\newtheorem{lemma}{Lemma}
\newtheorem{corollary}{Corollary}
\newtheorem{proposition}{Proposition}

\pgfplotscreateplotcyclelist{default}{%
	blue,mark=*\\%
	red,mark=star\\%
	teal,mark=square*\\%
	brown!60!black,mark=otimes*\\%
}

\newcolumntype{P}[1]{>{\centering\arraybackslash}p{#1}}

\begin{document}
\bstctlcite{IEEEexample:BSTcontrol}
\title{Globally Optimal Spectrum- and Energy-Efficient Beamforming for Rate Splitting Multiple Access}

\author{
	Bho~Matthiesen,~\IEEEmembership{Member,~IEEE},
	Yijie~Mao,~\IEEEmembership{Member,~IEEE},
	Armin~Dekorsy,~\IEEEmembership{Senior Member,~IEEE},
	Petar~Popovski,~\IEEEmembership{Fellow,~IEEE},
	Bruno~Clerckx,~\IEEEmembership{Fellow,~IEEE},
	%}
	\thanks{
		Parts of this paper were presented at 2021 IEEE International Conference on Acoustics, Speech and Signal Processing \cite{icassp2021}.
	}%
	\thanks{
	B.~Matthiesen and A.~Dekorsy are with the Department of Communications Engineering, University of Bremen, 28359 Bremen, Germany (e-mail: \{matthiesen, dekorsy\}@ant.uni-bremen.de).
	Y.~Mao is with School of Information Science and Technology, ShanghaiTech University, 387433 Shanghai, China, (e-mail: maoyj@shanghaitech.edu.cn).
	P.~Popovski is with the Department of Electronic Systems, Aalborg University, 9220 Aalborg, Denmark. He is also a U~Bremen Excellence Chair in the Department of Communications Engineering, University of Bremen, 28359 Bremen, Germany (e-mail: petarp@es.aau.dk).
	B. Clerckx is with the Department of Electrical and Electronic Engineering, Imperial College London, London SW7 2AZ, U.K. (e-mail: b.clerckx@imperial.ac.uk).
	}%
	\thanks{This work is supported in part by the German Research Foundation (DFG) under grant EXC 2077 (University Allowance), by the U.K. Engineering and Physical Sciences Research Council (EPSRC) under grants EP/N015312/1 and EP/R511547/1, and by the North-German Supercomputing Alliance (HLRN).
	}
}

\maketitle
%\tikzset{external/force remake}
%\tikzset{external/export=false}

\begin{abstract}
	Rate splitting multiple access (RSMA) is a promising non-orthogonal transmission strategy for next-generation wireless networks.
%%%
It has been shown to outperform existing multiple access schemes in terms of spectral and energy efficiency when suboptimal beamforming schemes are employed.   
%%%
In this work, we fill the gap between suboptimal and truly optimal beamforming schemes and conclusively establish the superior spectral and energy efficiency of RSMA. To this end, we propose a
successive incumbent transcending (SIT) branch and bound (BB) algorithm to find globally optimal beamforming solutions that maximize the weighted sum rate or energy efficiency of RSMA in Gaussian multiple-input single-output (MISO) broadcast channels.
%%%
Numerical results show that RSMA exhibits an explicit globally optimal spectral and energy efficiency gain over conventional multi-user linear precoding (MU-LP) and power-domain non-orthogonal multiple access (NOMA). 
%%%
Compared to existing globally optimal beamforming algorithms for MU-LP, the proposed SIT BB not only improves the numerical stability but also achieves faster convergence.  
%%%
Moreover, for the first time, we show that the spectral/energy efficiency of RSMA achieved by suboptimal beamforming schemes (including weighted minimum mean squared error (WMMSE) and successive convex approximation) almost coincides with the corresponding globally optimal performance, making it a valid choice for performance comparisons.
%%%
The globally optimal results provided in this work are imperative to the ongoing research on RSMA as they serve as benchmarks for  existing suboptimal beamforming strategies and those to be developed in multi-antenna broadcast channels.
\end{abstract}
\glsresetall
\begin{IEEEkeywords}
Rate splitting multiple access (RSMA), rate splitting, global optimization, spectral efficiency, energy efficiency, \cgls{miso}, \cgls{bc}, interference networks, next generation multiple access, non-orthogonal transmission 
\end{IEEEkeywords}
\glsresetall

\section{Introduction}
Over the past few years, \cgls{rsma}, built upon the concept of \cgls{rs},  has emerged as a promising non-orthogonal \cgls{phy}-layer transmission paradigm for interference management and multiple access in modern multi-antenna communication networks \cite{RSintro16bruno,mao2017eurasip,mao2019beyondDPC}. 
%%%
The key design principle of \cgls{rsma} is to partially decode the multi-user interference and partially treat it as noise. This is done by splitting user messages into common and private parts, respectively, and then transmitting them employing superposition coding \cite{mao2022survey}. 
%%%
The common message is decoded by multiple users, while the private message is only decoded by the corresponding user employing \cgls{sic}.
%%%
By flexibly adjusting the message splits according to the interference level, \cgls{rsma} allows arbitrary combinations of joint decoding and treating interference as noise. 
%%%
Therefore, \cgls{rsma} is a powerful interference management  and multiple access strategy that softly bridges and subsumes existing schemes such as \cgls{sdma}, which fully treats interference as noise,  \cgls{noma}, which fully decodes interference, and \cgls{oma}, which completely avoids interference by allocating orthogonal radio resources among users \cite{mao2017eurasip, bruno2019wcl}.
%%%

The concept of \cgls{rs} was introduced forty years ago for the two-user \cgls{siso} \cgls{ic} \cite{carleial1978RS,Han1981}. However, the use of \cgls{rs} as a building block of \cgls{rsma} is motivated by the recent progress of \cgls{rs}-based network design in multi-antenna wireless networks:  \cgls{rs} was first shown to achieve the optimal \cgls{dof} region  of  underloaded \cgls{miso} \cglspl{bc} with partial \cgls{csit} in \cite{enrico2017bruno} and of overloaded \cgls{miso} \cglspl{bc} with heterogeneous \cgls{csit} in \cite{mao2021IoT}. 
%%%
The \cgls{dof} benefits of \cgls{rs} subsequently motivated investigations of \cgls{rsma} precoder design at finite
\cglspl{snr} \cite{hamdi2016tcom,mao2017eurasip,mao2018EE,mao2019TCOM,bruno2019wcl,mao2019maxmin,Zheng2020JSAC,fuhao2020secrecyRS}. 
%%%
%%%
%
There are two general lines of research for \cgls{rsma} resource allocation, namely, low-complexity beamforming design \cite{RS2015bruno,onur2021mobility,Lu2018MMSERS,bruno2019wcl,Gui2020EESEtradeoff,Lu2018MMSERS,Minbo2016MassiveMIMO,Asos2018MultiRelay,bruno2020MUMIMO} and beamforming optimization \cite{hamdi2016tcom,mao2017eurasip,alaa2020cranimperfectCSIT,mao2021IoT,mao2019maxmin, fuhao2020secrecyRS, Zheng2020JSAC,Gui2020EESEtradeoff,mao2018EE,alaa2020EECRAN,onur2021DAC,RSswiptIC2019CL,Camana2020swiptRS,wonjae2021imperfectCSIR} mainly with respect to maximizing the \cgls{se} or \cgls{ee}. 
%%%
Low-complexity beamforming approaches such as using
\gls{rbf} \cite{RS2015bruno,enrico2016bruno,onur2021mobility,Lu2018MMSERS},
\gls{mbf} \cite{Minbo2016MassiveMIMO,AP2017bruno,Asos2018MultiRelay,Lu2018MMSERS,bruno2019wcl,Gui2020EESEtradeoff}, or \gls{svd} \cite{hamdi2016tcom}
for the common stream together with   \gls{zf} \cite{RS2015bruno,enrico2016bruno,onur2021mobility} or \gls{rzf}/\gls{mmse} \cite{Minbo2016MassiveMIMO,AP2017bruno,Asos2018MultiRelay,Lu2018MMSERS} for the private streams have been widely studied in \cgls{rsma}-aided \cgls{miso} \cgls{bc}. 
%%%
 \cGls{bd} \cite{bruno2020MUMIMO,onur2021mobility} has been further investigated when the receivers are equipped with multiple antennas.
Instead, precoder optimization strives to find optimal beamformers that maximize achievable  performance  regions of RSMA.
%%%
Existing beamforming design algorithms such as \cgls{wmmse} \cite{hamdi2016tcom,mao2017eurasip,alaa2020cranimperfectCSIT,mao2021IoT,mao2019TCOM}, \cgls{sca} \cite{mao2019maxmin, Zheng2020JSAC,mao2019TCOM,Gui2020EESEtradeoff,mao2018EE,alaa2020EECRAN,fuhao2020secrecyRS,wonjae2021imperfectCSIR,Jia2020SEEEtradeoff,alaa2020EECRAN}, \gls{admm} \cite{xu2021rate,onur2021DAC,rafael2021radarsensing}, and \gls{sdr} \cite{fuhao2020secrecyRS,RSswiptIC2019CL,Camana2020swiptRS,wonjae2021imperfectCSIR} have been investigated for RSMA.
%%%
\cgls{sca}-based algorithms follow the classical idea of successively approximating the original non-convex problem with a sequence of convex approximations.  
%%%
\cgls{wmmse} and \cgls{admm}-based algorithms are the block-wise alternative optimization where the variables are divided into  blocks and the original problem is optimized alternatively with respect to a single block of variables while the rest of the blocks are held fixed.
%%%
All of them could only guarantee a locally optimal point of the original problem \cite{razaviyayn2014SCA}.
%%%
Though \cgls{sdr} has the potential to find the global optimum, existing works all focus on combining SDR with other efficient approaches such as SCA \cite{fuhao2020secrecyRS,wonjae2021imperfectCSIR},  gradient-based approach \cite{RSswiptIC2019CL}, particle swarm optimization  \cite{Camana2020swiptRS},  heuristic approaches \cite{Medra2018SPAWC}  in order to reduce the computational complexity. 
%%%
Therefore, the solutions obtained in these works cannot ensure global optimality.
%%%
\Cref{tab:algosurvey}  summarizes the state-of-the-art beamforming design approaches that have been proposed for \cgls{rsma}.
%%%
None of them exhibits strong optimality guarantees. 
%%%
Hence, all performance analyses based on these approaches are incapable of conclusively establishing the superiority of \cgls{rsma} over the previously mentioned multiplexing schemes.
To the best of the authors knowledge, there is no existing work focusing on the globally optimal beamforming design of \cgls{rsma}, and the maximum \cgls{se} and \cgls{ee} performance achieved by RSMA remains unknown.
%%%

%%%
\begin{table}
\caption{Survey of existing \cgls{rsma}  beamforming design approaches.}
	\label{tab:algosurvey}
	\centering
	\footnotesize
	\begin{tabularx}{\linewidth}{@{ }ll@{\enskip}Xc@{ }}
		\toprule
		%%%%%%%%%%
		\multicolumn{2}{c}{Beamforming approach}
		& Maximize SE
		& Maximize EE
		\\
		\midrule
		Low
		& RBF + ZF
		& \cite{RS2015bruno,enrico2016bruno,onur2021mobility}
		& ---
		\\
		%%%%%%%%
		complexity
		& RBF + RZF
		& \cite{Lu2018MMSERS}
		& ---
		\\ 
		%%%%%%%
		& MBF + ZF
		& \cite{bruno2019wcl}
		& \cite{Gui2020EESEtradeoff}
		\\ 
		%%%%%%%%%%
		& MBF + RZF
		& \cite{Lu2018MMSERS,Minbo2016MassiveMIMO,AP2017bruno,Asos2018MultiRelay}
		& ---
		\\ 
		%%%%%%%%%%
		& SVD + ZF
		& \cite{hamdi2016tcom}
		& ---
		\\ 
		%%%%%%%%%%
		& MBF + BD
		& \cite{bruno2020MUMIMO}
		& ---
		\\
		\midrule
		%%%%%%%%
		%%%%%%%%
		Suboptimal
		& WMMSE-based
	& \cite{hamdi2016tcom,mao2017eurasip,alaa2020cranimperfectCSIT,mao2021IoT,mao2019TCOM}
		& ---
		\\ 
		%%%%%%%%%%
		& SCA-based
		& \cite{mao2019maxmin, Zheng2020JSAC,mao2019TCOM,Gui2020EESEtradeoff,fuhao2020secrecyRS,wonjae2021imperfectCSIR}
		& \cite{mao2018EE,alaa2020EECRAN,Jia2020SEEEtradeoff}         \\ 
		%%%%%%%%%%
		& ADMM-based
		& \cite{xu2021rate,onur2021DAC,rafael2021radarsensing}
		& ---
		\\ 
		%%%%%%%%%%
		& SDR-based
		& \cite{fuhao2020secrecyRS,wonjae2021imperfectCSIR}
		& ---
		\\
		\midrule
		%%%%%%%%%%
		Globally
		& SIT BB
		& This work
		& This work
		\\
		optimal
		\\
		\bottomrule
	\end{tabularx}
\end{table}
%%%

The goal of this paper is to bridge this gap and derive an algorithm to determine a globally optimal beamforming solution for \cgls{rsma} with respect to \cgls{wsr} and \cgls{ee} maximization.
The corresponding optimization problem is related to joint multicast and unicast precoding that is known to be NP-hard \cite{Sidiropoulos2006,Luo2008}. While several globally optimal algorithms for unicast beamforming \cite{Bjornson2013,Tervo2015} and multicast beamforming \cite{Lu2017} exist, joint solution methods are scarce. In particular, the procedure in \cite{Liu2017} solves the power minimization problem and \cite{Chen2018} maximizes the \cgls{wsr} for joint multicast and unicast beamforming.
All these methods are based on \cgls{bb} in combination with the \cgls{soc} transformation in \cite{Bengtsson1999}. However, as this transformation moves the complexity into the feasible set, pure \cgls{bb} methods are prone to numerical problems, see \cref{sec:sitfund}.
Instead, in this paper we design a \cgls{sit} \cgls{bb} algorithm to solve this beamforming problem with improved numerical stability and faster convergence. To the best of the authors knowledge, this is the first globally optimal solution algorithm for an instance of the joint unicast and multicast problem with respect to \cgls{ee} maximization.

To summarize, the contributions of this paper are:
\begin{enumerate}
	\item We develop a numerical solver for the \cgls{wsr} and \cgls{ee} beamforming problem in \cgls{rsma} with guaranteed convergence to a globally optimal solution. We emphasize the novelty of the globally optimal \cgls{ee} maximization method for an instance of the joint unicast and multicast beamforming problem.
	\item We apply the \acrfull{sit} principle to a \cgls{miso} beamforming problem. The proposed algorithm incorporates \cgls{mulp} and  2-user \cgls{noma} beamforming as special cases. It exhibits faster practical convergence and improved numerical stability over state-of-the-art solution methods for \cgls{mulp} beamforming.
	\item From a theoretical perspective, we establish finite convergence to the optimal solution. This property does not hold for most \cgls{bb}-based beamforming algorithms.
	\item Extensive numerical verification is done, both to assess the numerical properties of the proposed algorithm and to evaluate the performance of suboptimal state-of-the-art methods. In particular, we show that these methods, including \cgls{wmmse} and \cgls{sca}, are often close to the true optimum solution.
\end{enumerate}

The paper organization continues as follows. In the next section, we define the system model, formally state the optimization problem and transform it into an equivalent form more suitable for numerical solution. In \cref{sec:sitfund}, the mathematical fundamentals of the proposed algorithm are reviewed. These are applied in \cref{sec:alg} to derive the solution algorithm and prove its convergence. We close the paper with numerical experiments in \cref{sec:numeval} and a short discussion.% in \cref{sec:conclusions}.

\subsubsection*{Notation}
Scalars and functions are typeset in normal font $x$.
The absolute value of $\cdot$ is $|\cdot|$ and $v(\text{n})$ is the optimal value of the optimization problem in equation (n).
$\Re\{\cdot\}$ and $\Im\{\cdot\}$ are the real and imaginary parts of a complex number, $j$ is the imaginary unit, and $\angle\cdot$ is the argument of $(\cdot)$.
A vector $\vec x$ has components $[x_1, x_2, \dots]^T$ and is a column vector unless noted otherwise. The all-zero and all-ones vectors are denoted as $\vec 0$ and $\vec 1$, respectively. The operators
$(\cdot)^T$,
$(\cdot)^H$, and
$\Vert\cdot\Vert$
are the transpose, the conjugate transpose and the Euclidean norm, respectively. Scalar operators are applied element-wise to vectors, where relational operators evaluate to true if they hold element-wise for all elements.
A set is written as $\mathcal X$ and a family of sets as $\mathscr X$. The sets of real and complex numbers are denoted as $\mathds R$ and $\mathds C$.
The notation $\mathcal X\setminus x$ is a shorthand for $\mathcal X\setminus\{x\}$.
Let $\mathcal X$ and $(\vec x, \vec y)\in\mathcal X$. Then, $\proj_{\vec x} \mathcal X = \{ \vec x : (\vec x, \vec y)\in\mathcal X\ \text{for some}\ \vec y\}$, i.e., the projection of $\mathcal A$ onto the $\vec x$ coordinates.

\section{System Model \& Problem Statement}\label{sec:sysmod}

Consider the downlink in a wireless network where an $M$ antenna \cgls{bs} serves $K$ single-antenna users. The received signal at user $k$, $k\in\mathcal K = \{ 1, \dots, K \}$, for each channel use is
$y_k=\vec{h}_k^{H}\vec{x}+n_k$,
where the transmit signal $\vec x\in\mathds{C}^{M\times 1}$ is subject to an average power constraint $P$, $\vec h_k$ is the complex-valued channel from the \cgls{bs} to user $k$, and $n_k$ is circularly symmetric complex white Gaussian noise with unit power at user $k$. We assume perfect \cgls{csi} at the transmitter and receivers.

\begin{figure}
	\centering
	\includegraphics[width=3.1in]{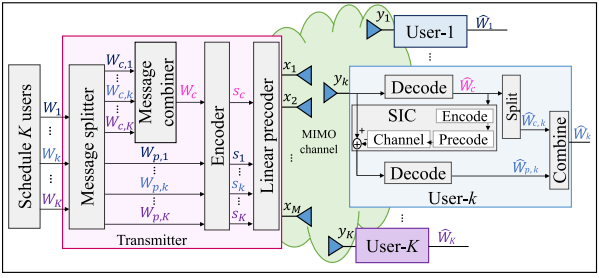}%
	\caption{1-layer RS model for $K$ users, where the common stream $s_c$ is shared by all users.}
	\label{fig: 1-layer RS transmission model}
\end{figure}

The transmitter employs 1-layer \cgls{rs} \cite{mao2017eurasip,hamdi2016tcom} as illustrated in \cref{fig: 1-layer RS transmission model}, i.e., it splits the message $W_k$ intended for user $k$ into a common part $W_{c,k}$ and a private part $W_{p,k}$. Then, the common messages $W_{c,1}, \ldots, W_{c,K}$ are combined into a single message $W_c$ and the $K+1$ resulting messages are encoded into independent Gaussian data streams $s_c, s_1, \dots, s_K$, each having unit power. These symbols are combined with linear precoding into the transmit signal
$\vec x = \vec p_c s_c + \sum_{k\in\mathcal K} \vec p_k s_k$,
where $\vec p_c, \vec p_1, \dots, \vec p_K \in \mathds C^M$ are the precoding vectors.
Due to the average transmit power constraint, they need to satisfy $\Vert\vec p_c\Vert^2 + \sum_{k\in\mathcal K} \Vert\vec p_k \Vert^2 \le P$.

Each receiver $k\in\mathcal K$ uses \cgls{sic} to first recover $s_c$ and then $s_k$ from its received signal $y_k$.
In particular, $s_c$ is decoded first by treating interference from all other streams as noise. This allows user $k$ to recover its desired common part $W_{c,k}$.
%%%
Then,  $s_c$  is cancelled from the received signal and user $k$ proceeds to decode $s_k$ to recover the desired private part $W_{p,k}$. These two messages are combined to obtain $W_k$.

Given a precoding scheme $\vec p_c, \vec p_1, \dots, \vec p_K$, asymptotic error free decoding of $W_c$ and $W_{p,k}$ is possible if the rates of these messages satisfy 
\begin{align}
	R_c &\le \log(1 + \gamma_{c,k}), \forall k\in\mathcal{K},
	&
	R_{p,k} &\le \log(1+\gamma_{p,k})
	\label{eq: instantanesous}
\end{align}
with \cglspl{sinr}
\begin{align} \label{eq: SINR cp}
	\gamma_{c,k}&=\frac{|{\vec{h}}_{k}^{H}\vec{p}_{c}|^2}{\sum\limits_{j\in\mathcal{K}}|\vec{h}_{k}^{H}\vec{p}_{j}|^2+1},
	&
	\gamma_{p,k}&=\frac{|{\vec{h}}_{k}^{H}\vec{p}_{k}|^2}{\sum\limits_{j\in\mathcal{K}\setminus k}|\vec{h}_{k}^{H}\vec{p}_{j}|^2+1}.
\end{align}
The rate $R_{c} $ is shared across the users, where
user $k$  is allocated a portion ${C}_{k}$ corresponding to the rate of $W_{c,k}$,
such that $\sum_{k \in \mathcal{K}} {C}_{k} = {R}_{c} $. The total rate of user $k$ is $R_k = C_k + R_{p,k}$.

Observe that this system model can be interpreted as an instance of the joint multicast and unicast beamforming problem. It includes several notable special cases. With $\vec p_c = 0$, we obtain \cgls{mulp} and for $\vec p_k = 0$, $k \in \mathcal K$, it is the multicast beamforming problem. It also includes 2-user \cgls{noma} \cite{bruno2019wcl}.

\subsection{Problem Statement} \label{sec:transf}
We consider \cgls{wsr} maximization under minimum rate \cgls{qos} constraints, i.e.,
\begin{subequations}
	\label{eq:WSR QoS prob}
	\begin{align}
		\max_{\substack{\vec p_1, \dots, \vec p_K, \vec p_c,\\ \vec C, \vec R_c, \vec R_p, \vec \gamma_c, \vec \gamma_p}}\quad
		& \sum_{k\in\mathcal{K}} u_k \left(C_k+R_{p,k}\right)\\
	\mbox{s.t.}\quad
& R_{c}, R_{p,k}, \gamma_{c,k} \text{ and } \gamma_{p,k} \text{ as in \eqref{eq: instantanesous}--\eqref{eq: SINR cp}}\label{eq:WSR QoS prob:2} \\
	& \sum_{k'\in \mathcal{K}}C_{k'}\leq R_{c} \label{eq:WSR QoS prob:3} \\
	& C_k\geq 0,\enskip k\in\mathcal{K} \label{eq:WSR QoS prob:4} \\
	& C_k+R_{p,k}\geq R_k^{th},\enskip k\in\mathcal{K} \label{eq:WSR QoS prob:6} \\
	& \Vert \vec p_c \Vert^2 + \sum_{k\in\mathcal K} \Vert \vec p_k \Vert^2 \le P \label{eq:WSR QoS prob:5}
	\end{align}
\end{subequations}
with nonnegative weight vector $\vec{u}=[u_1,\ldots,u_K]^T \neq \vec 0$, where \cref{eq:WSR QoS prob:2} includes the rate constraints and \cgls{sinr} definition from the system model, \cref{eq:WSR QoS prob:2} ensures that the common rate allocations $C_k$ are feasible, \cref{eq:WSR QoS prob:3} implements the non-negativity of the common rates, \cref{eq:WSR QoS prob:6} is the minimum rate constraint with threshold $R_k^{th}$ for user $k$, and \cref{eq:WSR QoS prob:5} is the maximum power constraint at the \cgls{bs}.
We also consider
\cgls{ee} maximization
\begin{subequations}
	\label{eq:EE prob}
	\begin{align}
		\max_{\substack{\vec p_1, \dots, \vec p_K, \vec p_c,\\ \vec C, R_c, \vec R_p, \vec \gamma_c, \vec \gamma_p}}\quad
		& \frac{\sum_{k\in\mathcal{K}} C_k+R_{p,k}}{\mu \left( \Vert \vec p_c \Vert^2 + \sum_{k\in\mathcal K} \Vert \vec p_k \Vert^2 \right) + P_c} \\
	\mbox{s.t.}\quad
	& \text{\cref{eq:WSR QoS prob:2,eq:WSR QoS prob:3,eq:WSR QoS prob:4,eq:WSR QoS prob:5,eq:WSR QoS prob:6}},
	\end{align}
\end{subequations}
where $\mu \ge 0$ is the power amplifier inefficiency and $P_c > 0$ is the static circuit power consumption.
For notational simplicity, we define $\vec C = [C_1, \dots ,C_k]^T$, $\vec \gamma_{p} = [\gamma_{1}, \dots, \gamma_K]^T$, $\vec \gamma_{c} = [\gamma_{c,1}, \dots, \gamma_{c,K}]^T$, $\vec R_p = [R_{p,1}, \dots, R_{p,K}]^T$.
Both problems can be combined into the equivalent optimization problem
\begin{subequations} \label{eq:srmaxequiv}
	\begin{align}
		\max_{\substack{\vec p_1, \dots, \vec p_K,\\ \vec p_c, \vec C, \vec \gamma_c, \vec \gamma_p}}\quad
		&\frac{\sum_{k\in\mathcal{K}} u_k \left(C_k+\log(1+ \gamma_{p,k})\right)}{\mu \left( \Vert \vec p_c \Vert^2 + \sum_{k\in\mathcal K} \Vert \vec p_k \Vert^2 \right) + P_c}\\
	\mbox{s.t.}\quad
	& \gamma_{c,k} \text{ and } \gamma_{p,k} \text{ as in \eqref{eq: SINR cp}}\label{eq:srmaxequiv:2} \\
	& \sum_{k'\in \mathcal{K}}C_{k'}\leq \log(1+ \gamma_{c,k}), \enskip k\in\mathcal{K} \label{eq:srmaxequiv:3} \\
	& C_k \geq \max\!\left\{ 0, R_k^{th}\! - \log(1+ \gamma_{p,k}) \right\}\!,  k\in\mathcal{K} \label{eq:srmaxequiv:4} \\
	& \Vert \vec p_c \Vert^2 + \sum_{k\in\mathcal K} \Vert \vec p_k \Vert^2 \le P \label{eq:srmaxequiv:6},
	\end{align}
\end{subequations}
where \cref{eq:srmaxequiv:4} combines \cref{eq:WSR QoS prob:4,eq:WSR QoS prob:6} into a single constraint.
Clearly, we obtain \cgls{wsr} maximization for $\mu = 0, P_c = 1$ and \cgls{ee} maximization for $\vec u = \vec 1$.

A globally optimal solution of \cref{eq:srmaxequiv} can be obtained by solving
\begin{subequations} \label{eq:srmax}
	\begin{align}
		\max_{\substack{\vec p_c, \vec p_1, \dots, \vec p_K,\\ \vec C, \vec \gamma_p, s, \vec d, \vec e}}{} &  \frac{\sum_{k\in\mathcal{K}} u_k \left(C_k+\log(1+ \gamma_{p,k})\right)}{\mu \left( \Vert \vec p_c \Vert^2 + \sum_{k\in\mathcal K} \Vert \vec p_k \Vert^2 \right) + P_c}\\
	\mbox{s.t.}\quad
	& \sqrt{\gamma_{p,k}} \smash{\left( \sum\nolimits_{j\in\mathcal{K}\setminus k}|\vec{h}_{k}^{H}\vec{p}_{j}|^2+1 \right)^{1/2}}\!\!\! \le {\vec{h}}_{k}^{H}\vec{p}_{k}  \label{eq:srmax:1} \\
	& \sqrt s \left( \sum\nolimits_{j\in\mathcal{K}}|\vec{h}_{1}^{H}\vec{p}_{j}|^2+1 \right)^{1/2} \le {\vec{h}}_{1}^{H}\vec{p}_{c}  \label{eq:srmax:2} \\
	& \sqrt s \left( \sum\nolimits_{j\in\mathcal{K}}|\vec{h}_{k}^{H}\vec{p}_{j}|^2+1 \right)^{1/2}\! \le d_k, \forall k > 1 \label{eq:srmax:3} \\
	& (e_k, d_k) \in \mathcal C, \forall k > 1 \label{eq:srmax:8} \\
	& \Re\{\vec h_k^H \vec p_k\} \ge 0,\quad \Im\{\vec h_k^H \vec p_k\} = 0 \label{eq:srmax:4} \\
	& \Re\{\vec h_1^H \vec p_c\} \ge 0,\quad \Im\{\vec h_1^H \vec p_c\} = 0 \label{eq:srmax:5} \\
	& \forall k > 1: d_k \ge 0,\enskip e_k = {\vec{h}}_{k}^{H}\vec{p}_{c}\label{eq:srmax:7} \\
	& \sum_{k\in \mathcal{K}}C_{k}\leq \log(1+ s) \label{eq:srmax:9} \\
	& \eqref{eq:srmaxequiv:4} \text{ and } \eqref{eq:srmaxequiv:6} \label{eq:srmax:10}
	\end{align}
\end{subequations}
with
\begin{equation} \label{eq:circleset}
	(e, d) \in \mathcal C = \{ e\in\mathds C, d \in \mathds R : d  \le | e | \}
\end{equation}
instead of \cref{eq:srmaxequiv}.
A crucial observation is that this problem is a \cgls{socp} for fixed $s$, $\vec\gamma_p$ except for constraint \cref{eq:srmax:7}. Hence, the nonconvexity of \cref{eq:srmaxequiv} is only due to the \cgls{sinr} expressions and not due to the beamforming vectors. We will exploit this partial convexity in the final algorithm to limit the numerical complexity.

The transformation from \cref{eq:srmaxequiv} to \cref{eq:srmax} relies on the observation that, for all other variables except the beamforming vectors fixed, the nonconvexity in \cref{eq:srmaxequiv} stemming from $\vec p_k$ is only due to the product $|\vec h_k^H \vec p_k|^2$. This can be seen from considering \cref{eq:srmaxinter:1} below and multiplying it with the denominator of its \cgls{rhs}. Then, for fixed $\gamma_{p,k}$, one almost obtains the \cgls{soc} \cref{eq:srmax:1}, except for $|\vec h_k^H \vec p_k|^2$. Further observing that the solution of \cref{eq:srmax} is rotationally invariant in $\vec p_k$, one can select the solution that results in $|\vec h_k^H \vec p_k|$ being nonnegative and real valued. This is achieved by introducing constraint \cref{eq:srmax:4} \cite{Bengtsson1999}.

Unfortunately, this transformation is not sufficient to eliminate $\vec p_c$ as a nonconvex variable. This is because there is not a single product $|\vec h_k^H \vec p_c|^2$ that needs to be made real but $K$, one for each user. Our approach is to first replace $\gamma_{c,1}, \dots, \gamma_{c,K}$ by a single nonconvex variable $s = \min_k \gamma_{c,k}$ and then apply the same transformation as for the precoders $\vec p_1, \dots, \vec p_K$ for \emph{only one} of the products $|\vec h_k^H \vec p_c|^2$ (here for $k = 1$). This results in the constraints \cref{eq:srmax:2,eq:srmax:5}. The remaining $K-1$ ``almost''-\cgls{soc}-constraints are then addressed using the argument cuts approach from \cite{Lu2017}. In particular, real valued auxiliary variables $d_k$ are introduced together with the relaxation that these must be within the circle $|\vec h_k^H \vec p_c|$. These considerations will be made rigorous in \cref{prop:srmax}.

Another equivalent variant of \cref{eq:srmaxequiv} that is interesting in its own right is
\begin{subequations} \label{eq:srmaxinter}
	\begin{align}
		\max_{\substack{\vec p_c, \vec p_1, \dots, \vec p_K,\\ \vec C, \vec \gamma_p, s}}\quad &  \frac{\sum_{k\in\mathcal{K}} u_k \left(C_k+\log(1+ \gamma_{p,k})\right)}{\mu \left( \Vert \vec p_c \Vert^2 + \sum_{k\in\mathcal K} \Vert \vec p_k \Vert^2 \right) + P_c} \\
	\mbox{s.t.}\quad
	& \gamma_{p,k}\le \frac{|{\vec{h}}_{k}^{H}\vec{p}_{k}|^2}{\sum_{j\in\mathcal{K}\setminus k}|\vec{h}_{k}^{H}\vec{p}_{j}|^2+1} \label{eq:srmaxinter:1} \\
	& s \le \frac{|{\vec{h}}_{k}^{H}\vec{p}_{c}|^2}{\sum_{j\in\mathcal{K}}|\vec{h}_{k}^{H}\vec{p}_{j}|^2+1},\enskip k\in\mathcal K \label{eq:srmaxinter:2} \\
	& \eqref{eq:srmaxequiv:4}, \eqref{eq:srmaxequiv:6}  \text{ and } \eqref{eq:srmax:9}.
	\end{align}
\end{subequations}
This problem is obtained as a side product when showing the equivalence of \cref{eq:srmaxequiv} and \cref{eq:srmax}, which is established next.
\begin{proposition} \label{prop:srmax}
	Let $\vec x^\star = (\vec p_c^\star, \vec p_1^\star, \dots, \vec p_K^\star, \vec C^\star, \vec \gamma_p^\star)$.
	A point $(\vec x^\star, s^\star)$ solves \cref{eq:srmaxinter} if $(\vec x^\star, \vec \gamma_c^\star)$ solves \cref{eq:srmaxequiv} and $s^\star = \min_k \gamma_{c,k}^\star$.
	Conversely, the point $(\vec x^\star, \vec \gamma_c^\star)$ solves \cref{eq:srmaxequiv} if $(\vec x^\star, s^\star)$ solves \cref{eq:srmaxinter} and
	$\gamma_{c,k}^\star = \frac{|{\vec{h}}_{k}^{H}\vec{p}_{c}^\star|^2}{\sum_{j\in\mathcal{K}}|\vec{h}_{k}^{H}\vec{p}_{j}^\star|^2+1}$ for all $k\in\mathcal K$.
	Moreover, if $(\vec x^\star, s^\star, \vec d^\star, \vec e^\star)$ solves \cref{eq:srmax}, then $(\vec x^\star, s^\star)$ solves \cref{eq:srmaxinter} and $(\vec x^\star, \vec \gamma_c^\star)$ solves \cref{eq:srmaxequiv}, where $s^\star$ and $\vec\gamma_c^\star$ are as before.
\end{proposition}

\begin{IEEEproof}
	See Appendix~\ref{proof:prop:srmax}.
\end{IEEEproof}

\begin{corollary}
	Problems~\cref{eq:srmaxequiv,eq:srmax,eq:srmaxinter} have the same optimal value.
\end{corollary}

In the next section, we introduce some mathematical preliminaries before we develop a solution algorithm for \cref{eq:srmax} in \cref{sec:alg}.

\section{Mathematical Background} \label{sec:sitfund}
Problem~\cref{eq:srmax} is an NP-hard nonconvex optimization problem. To see this, consider problem \cref{eq:srmaxinter} for $\mu = 0$, $u_k = 1$, and fix all variables except $\vec p_c$, $\vec C$ and $s$. Then, it is equivalent to
\begin{equation}
	\max_{\vec p_c} \min_k \{ | \tilde{\vec h}_k^H \vec p_c |^2 \} \quad\st\quad \Vert \vec p_c \Vert^2 \le \tilde P.
\end{equation}
This is known as multicast beamforming and shown to be NP-hard in \cite{Sidiropoulos2006}. Our solution approach for this part of the problem relies on the so-called ``argument cuts'' proposed in \cite{Lu2017}. The introduction of the auxiliary variables $d_k$ and $e_k$ in \cref{eq:srmax} is motivated by this approach and collects most of the nonconvexity due to $\vec p_c$ in \cref{eq:srmax:8}. The idea is to add a box constraint on $\arg(e_k)$ to $\mathcal C$ and then optimize over its convex envelope to obtain a bound on \cref{eq:srmax} suitable for a \cgls{bb} procedure.

The remaining nonconvexity in \cref{eq:srmax} stems from $\gamma_{p,k}$ and $s$ in \cref{eq:srmax:1,eq:srmax:2,eq:srmax:3}.  Previous global optimization algorithms for such problems rely on \cgls{bb} procedures with \cgls{socp} bounding \cite{Bjornson2013,Tervo2015,Liu2017,Chen2018}. However, this leads to an infinite algorithm where the convergence to the global optimal solution cannot be guaranteed in a finite number of iterations.\footnote{While this often does not lead to problems in practice, slower convergence might be observed in infinite algorithms. In addition, finiteness is an important theoretical aspect that differentiates a ``computational method'' from an ``algorithm'' \cite{Knuth1997vol1}.}
This is because the difficulty in solving \cref{eq:srmax} is due to the feasible set, while \cgls{bb} works best if the nonconvexity is mostly due to the objective. Please refer to \cite{mmp,sit,diss} for a detailed discussion of this topic. For the solution of \cref{eq:srmax}, finite convergence can be obtained by adding a line search procedure to every iteration of the \cgls{bb} procedure that recovers a feasible point and requires the solution of several \cgls{soc} feasibility problems \cite[Alg.~3]{Bjornson2013}. Hence, finite convergence in \cgls{bb} procedures comes at the cost of increased computational complexity.
Moreover, the auxiliary \cgls{socp} that is solved in every iteration of the \cgls{bb} procedure is numerically challenging as the feasible set can become very small. This leads to numerical problems even with commercial state-of-the-art solvers like Mosek \cite{mosek}. A computationally more tractable modification is proposed in \cite[\S 2.2.2]{Bjornson2013} that comes at the price of much harder feasible point acquisition in the \cgls{bb} procedure.

Instead, we design an algorithm based on the \cgls{sit} scheme \cite{Tuy2005a,Tuy2009,sit,diss} and combine it with a \cgls{brb} procedure. The resulting algorithm is numerically stable, has proven finite convergence, solves \cgls{ee} maximization and is the first global optimization algorithm specifically designed for \cgls{rsma}. Practically, it outperforms algorithms for similar problems as will be verified in \cref{sec:numeval}.
To better illustrate the core principles of \cgls{sit}, we first consider the following general optimization problem
\begin{equation} \label{eq:genopt}
	\max_{(\vec x, \vec \xi)\in\mathcal D}\enskip f(\vec x, \vec\xi) \quad\st\quad g_i(\vec x, \vec\xi) \le 0,\enskip i = 1, \dots, n
\end{equation}
with continuous, real-valued functions $f, g_1, \dots, g_n$ and nonempty feasible set. Further, assume that $f$ is concave,\footnote{Although this assumption does not hold for \cref{eq:srmax}, it will be established later that the \cgls{sit} approach is still applicable. This is because the sole purpose of this convexity assumption is to obtain a convex feasible set in \cref{eq:gendual}.} $g_1, \dots, g_n$ are convex in $\vec\xi$ for fixed $\vec x$, and $\mathcal D$ is a closed convex set. Depending on the structure of $g_1, \dots, g_n$ in $\vec x$ this problem might be quite hard to solve for \cgls{bb} methods \cite{Tuy2016,sit}.\footnote{This is also true for outer approximation methods like the Polyblock algorithm \cite{Tuy2016}.}
By exchanging the objective and constraints of \cref{eq:genopt}, we obtain the so-called \cgls{sit} dual
\begin{equation} \label{eq:gendual}
	\min_{(\vec x, \vec\xi)\in\mathcal D}\enskip \max_i\{ g_i(\vec x, \vec\xi) \} \quad\st\quad f(\vec x, \vec\xi)  \ge \delta.
\end{equation}
Observe that the optimal value of \cref{eq:genopt} is greater than or equal to $\delta$ if the optimal value of \cref{eq:gendual} is less than or equal to zero.
Conversely, if the optimal value of \cref{eq:gendual} is greater than zero, the optimal value of \cref{eq:genopt} is less than $\delta$.
Hence, the optimal solution of \cref{eq:genopt} can be obtained by solving a sequence of \cref{eq:gendual} with increasing $\delta$.
Since the feasible set of \cref{eq:gendual} is closed and convex, it can be solved much easier by \cgls{bb} than \cref{eq:genopt}.

Obtaining the exact optimal solution to continuous real-valued optimization problems is often computationally infeasible, even for linear or convex problems. A widely employed practice is to accept any feasible point with objective value within a prescribed tolerance $\eta$ of the exact optimal value as a solution.
That is, a point $(\bar{\vec x}, \bar{\vec\xi})$
is called an $\eta$-optimal solution of \cref{eq:genopt} if, for all feasible points $(\vec x, \vec\xi)$,
\begin{equation} \label{eq:etaopt}
	f(\bar{\vec x}, \bar{\vec\xi}) \ge f(\vec x, \vec\xi) - \eta.
\end{equation}
Likewise, the constraints in \cref{eq:genopt} can be hard to satisfy numerically. The most common approach is to relax them by $\varepsilon$. However, for nonconvex feasible sets this can lead to completely wrong solutions \cite{Tuy2005a,Tuy2016,sit}. Instead, for the \cgls{sit} scheme, the constraints are tightened by $\varepsilon$, i.e., the problem to be solved is
\begin{equation} \label{eq:genoptepsilon}
	\max_{(\vec x, \vec\xi)\in\mathcal D}\enskip f(\vec x, \vec\xi) \quad\st\quad g_i(\vec x, \vec\xi) \le -\varepsilon,\enskip i = 1, \dots, n
\end{equation}
for some $\varepsilon > 0$. Any point in this feasible set is denoted as $\varepsilon$-essential feasible and a solution of this problem satisfying \cref{eq:etaopt} is called essential $(\varepsilon, \eta)$-optimal solution of \cref{eq:genopt}. This constraint tightening removes numerically instable points from the feasible set and is necessary to ensure finite convergence of the \cgls{sit} scheme.

The outlined duality between \cref{eq:genopt,eq:gendual} is formalized in the following lemma.
\begin{lemma} \label{lem:duality}
	For every $\varepsilon>0$, the $\varepsilon$-essential optimal value of \cref{eq:genopt} is less than $\delta$ if and only of the optimal value of \cref{eq:gendual} is greater than or equal to $-\varepsilon$.
\end{lemma}
\begin{IEEEproof}
	Direct consequence of \cite[Prop.~1]{Tuy2005a}.
	% if follows from the proof of (ii), only if from (i)
	% alternative: See \cite[Prop.~7.13]{Tuy2016} and the proof of \cite[Prop.~7.14]{Tuy2016}.
	% note that this is not a contradiction to the text below eq:gendual: there, I'm talking about the optimal value, here it's the essential optimal value
\end{IEEEproof}
We refer to \cref{eq:genopt} as the primal problem and \cref{eq:gendual} as the dual problem.\footnote{In this paper, the concept of duality is used with respect to the \cgls{sit} dual as discussed in this section and not in terms of Lagrange duality theory.}

\subsection{Successive Incumbent Transcending Algorithm} \label{sec:sit}
The discussion above leads to the \cgls{sit} algorithm as stated in \cref{alg:sit}.
The core problem is \ref{alg:sit:step1}, which is implemented by solving \cref{eq:gendual} with a modified rectangular \cgls{bb} procedure.
Such a procedure has exponential computational complexity in the number of optimization variables. Since \cref{eq:gendual} is a convex optimization problem for fixed $\vec x$, the \cgls{sit} \cgls{bb} procedure should only operate on the nonconvex variables $\vec x$ and employ a convex solver for $\vec\xi$ to limit the computational complexity.

\begin{algorithm}[tbh]
	\caption[\Acrshort{sit} Algorithm]{\Acrshort{sit} Algorithm \cite[\S 7.5.1]{Tuy2016}.}\label{alg:sit}
	\small
	\centering
	\begin{minipage}{\linewidth-.85em}
		\begin{enumerate}[label=\textbf{Step \arabic*},ref=Step~\arabic*,start=0,leftmargin=*]
			\item \label{alg:sit:step0} Initialize $(\bar{\vec x}, \bar{\vec\xi})$ with the best known nonisolated feasible solution and set $\delta = f(\bar{\vec x}, \bar{\vec\xi}) + \eta$; otherwise do not set $(\bar{\vec x}, \bar{\vec\xi})$ and choose  $\delta \le f(\vec x, \vec\xi)$ $\forall (\vec x, \vec\xi) \in \mathcal D$.
			\item\label{alg:sit:step1} Check if \cref{eq:genopt} has a nonisolated feasible solution $(\vec x, \vec\xi)$ satisfying $f(\vec x, \vec\xi) \ge \delta$; otherwise, establish that no such $\varepsilon$-essential feasible $(\vec x, \vec\xi)$ exists and go to \ref{alg:sit:step3}.
			\item Update $(\bar{\vec x}, \bar{\vec\xi}) \gets (\vec x, \vec\xi)$ and $\delta \gets f(\bar{\vec x}, \bar{\vec\xi}) + \eta$. Go to \ref{alg:sit:step1}.
			\item\label{alg:sit:step3} Terminate: If  $(\bar{\vec x}, \bar{\vec\xi})$ is set, it is an essential $(\varepsilon, \eta)$-optimal solution; else Problem~\eqref{eq:genopt} is $\varepsilon$-essential infeasible.
		\end{enumerate}
	\end{minipage}
\end{algorithm}

The general idea of \cgls{bb} is to relax the feasible set and then subsequently partition this relaxed set in such a way that upper and lower bounds on the objective value in each partition can be computed efficiently. As the partition is successively refined, these bounds approach each other until the optimal value is found.
For a rectangular \cgls{bb} procedure, the feasible set is relaxed into an initial box
\begin{equation}
	\mathcal M_0 = [\vec r^0, \vec s^0] = \{ \vec x : r^0_i \le x_i \le s^0_i \}
\end{equation}
satisfying $\mathcal M_0\supseteq \proj_{\vec x} \mathcal D$.
Further, a bounding function $\beta(\mathcal M)$, $\mathcal M\subseteq\mathcal M_0$, with
$\beta(\mathcal M) = \infty$ if $\proj_{\vec x} \mathcal F\cap\mathcal M = \emptyset$ and
\begin{equation} \label{eq:defbeta}
	\beta(\mathcal M) \le \min_{\substack{(\vec x, \vec\xi)\in\mathcal F, \vec x\in\mathcal M}} \max_i \{ g_i(\vec x, \vec\xi) \},
\end{equation}
otherwise is required,
where $\mathcal F =  \{ (\vec x, \vec\xi)\in \mathcal D : f(\vec x, \vec\xi) \ge \delta \}$ is the feasible set of \cref{eq:gendual}.
The algorithm subsequently partitions the relaxed feasible set $\mathcal M_0$ into smaller boxes and stores the current partition of $\mathcal M_0$ in a set $\mathscr R_k$. In iteration $k$, the algorithm uses best-first selection to determine the next branch, i.e.,
\begin{equation} \label{eq:selection}
	\mathcal M_k \in \argmin\{\beta(\mathcal M) \,|\, \mathcal M\in\mathscr R_k\},
\end{equation}
and then replaces $\mathcal M_k = [\vec r^k, \vec s^k]$ by two new subrectangles
\begin{subequations}
\label{eq:partition}
\begin{align}
	\mathcal M^- &= \{ \vec x  : r_j \le x_j \le v_j,\ r_i \le x_i \le s_i\ (i\neq j) \} \\
	\mathcal M^+ &= \{ \vec x : v_j \le x_j \le s_j,\ r_i \le x_i \le s_i\ (i\neq j) \}
\end{align}
\end{subequations}
with $\vec v = \frac{1}{2} (\vec s + \vec r)$ and $j \in\argmax_j s_j - r_j$. For each of these new boxes, a lower bound on the objective value is computed using the bounding function $\beta(\mathcal M)$. To ensure convergence, the bounding needs to be consistent with branching, i.e., $\beta(\mathcal M)$ has to satisfy
\begin{equation} \label{eq:consistent}
	\beta(\mathcal M) - \min_{\substack{(\vec x, \vec\xi)\in\mathcal F,\\ \vec x\in\mathcal M}} \max_i \{ g_i(\vec x, \vec\xi) \}
	\rightarrow 0 \enskip\mathrm{as}\enskip \max_{\vec x, \vec y\in\mathcal M} \Vert \vec x - \vec y \Vert \rightarrow 0,
\end{equation}
and a dual feasible point $\vec x^k\in\proj_{\vec x}\mathcal F\cap\mathcal M_k$ is required if $\beta(\mathcal M_k) < \infty$.
Suitable pruning and termination rules that ensure convergence can be obtained from the following lemma that is adapted from \cite[Prop.~7.14]{Tuy2016} and \cite[Prop.~5.9]{diss}.

\begin{lemma} \label{lem:conv}
	Let $\varepsilon> 0$ be given and define $g(\vec x, \vec\xi) = \max_i\{g_i(\vec x, \vec\xi)\}$. Let $\beta(\mathcal M)$ satisfy \cref{eq:defbeta,eq:consistent} and $\mathcal M_k$ be as in \cref{eq:selection}. Then, as $\max_{\vec x, \vec y\in\mathcal M_k} \Vert \vec x - \vec y \Vert \rightarrow 0$ for $k\to\infty$,
	either $g(\vec x^k, \vec\xi^*) < 0$ for some $k$ and $(\vec x^k, \vec\xi^*)\in\mathcal F$ or $\beta(\mathcal M_k) > -\varepsilon$ for some $k$. In the former case, $(\vec x^k, \vec\xi^*)$ is a nonisolated feasible solution of \cref{eq:genopt} satisfying $f(\vec x^k, \vec\xi^*)\ge\delta$. In the latter case, no $\varepsilon$-essential feasible solution $(\vec x, \vec\xi)$ of \cref{eq:genopt} exists such that $f(\vec x, \vec\xi)\ge\delta$.
\end{lemma}
\begin{IEEEproof}
	Please refer to \cite[Prop.~5.9]{diss}.
\end{IEEEproof}

This suggests a \cgls{bb} procedure with pruning criterion $\beta(\mathcal M) < -\varepsilon$ and termination criterion
\begin{equation} \label{eq:sit:primalfeas}
	 0 > \min\nolimits_{\vec\xi}\; g(\vec x^k, \vec\xi) \enskip\st\enskip (\vec x^k, \vec\xi)\in\mathcal F.
\end{equation}
In the following section, we apply this approach to find the solution of \cref{eq:srmax} and explicitly incorporate the outlined \cgls{bb} procedure into \cref{alg:sit}.

\section{Globally Optimal Beamforming} \label{sec:alg}
We design a globally optimal solution algorithm for \cref{eq:srmax} based on the fundamentals in the previous section.
There, we have seen that the \cgls{sit} algorithm requires a bounding function, a feasible point in each iteration and an initial box $\mathcal M_0$. These aspects will be discussed after identifying and discussing the \cgls{sit} dual. At the end of this section, we state the complete algorithm and establish its convergence.
We also derive a reduction procedure in \cref{sec:sit:red} that is essential for practical convergence, although it is not strictly necessary from a theoretical perspective.

The \cgls{sit} dual should contain all of the problem's nonconvexity in the objective function. Following the discussion in \cref{sec:transf}, the nonconvexity in \cref{eq:srmax} is due to \cref{eq:srmax:1,eq:srmax:2,eq:srmax:3,eq:srmax:8} and we obtain the \cgls{sit} dual as
\begin{subequations} \label{eq:sitdual}
	\begin{align}
		\min_{\substack{\vec p_c, \vec p_1, \dots, \vec p_K,\\ \vec C, \vec \gamma_p, s, \vec d, \vec e}} & \max\bigg[
			\sqrt s \left( \sum\nolimits_{j\in\mathcal{K}}|\vec{h}_{1}^{H}\vec{p}_{j}|^2+1 \right)^{1/2} - {\vec{h}}_{1}^{H}\vec{p}_{c}, \notag\\
			&\quad \max_{k > 1} \bigg\{\sqrt s \left( \sum\nolimits_{j\in\mathcal{K}}|\vec{h}_{k}^{H}\vec{p}_{j}|^2+1 \right)^{1/2}\!\!\!\!\! - d_k \bigg\}, \notag\\
			&\max_{k\in\mathcal K} \bigg\{ \sqrt{\gamma_{p,k}} \left( \sum\nolimits_{j\in\mathcal{K}\setminus k}|\vec{h}_{k}^{H}\vec{p}_{j}|^2+1 \right)^{1/2}\!\!\!\!\! - {\vec{h}}_{k}^{H}\vec{p}_{k} \bigg\}, \notag\\
			&\quad\max_{k > 1} \big\{d_k - |e_k| \big\} \bigg]\label{eq:sitdual:0}\\
	\mbox{s.t.}\quad
	&   \frac{\sum_{k\in\mathcal{K}} u_k \left(C_k+\log(1+ \gamma_{p,k})\right)}{\mu \left( \Vert \vec p_c \Vert^2 + \sum_{k\in\mathcal K} \Vert \vec p_k \Vert^2 \right) + P_c}\ge \delta  \label{eq:sitdual:1} \\
	& \text{\cref{eq:srmax:4,eq:srmax:5,eq:srmax:7,eq:srmax:9,eq:srmaxequiv:4,eq:srmaxequiv:6}}. \label{eq:sitdual:2}
	\end{align}
\end{subequations}
Observe that \cref{eq:sitdual:1} is equivalent to the \cgls{soc}
\begin{multline} \label{eq:sitdual:soc}
	\sum_{k\in\mathcal{K}} u_k \left(C_k+\log(1+ \gamma_{p,k})\right) \\\ge \delta \bigg( \mu \bigg( \Vert \vec p_c \Vert^2 + \sum_{k\in\mathcal K} \Vert \vec p_k \Vert^2 \bigg) + P_c \bigg)
\end{multline}
since the denominator in \cref{eq:sitdual:1} is positive. First, smoothen the objective by using the epigraph form with auxiliary variable $t$, and successively convert the pointwise maximum expressions to smooth constraints. Then, the new constraints $d_k - |e_k| \le t$, for $k > 1$, are equivalent to $(e_k, d_k - t)\in\mathcal C$. Introducing auxiliary variables $\alpha_k \in[0, 2\pi]$ and constraints $\alpha_k = \angle e_k$ for $k > 1$ leads to the equivalent optimization problem
\begin{subequations} \label{eq:sitdual2}
	\begin{align}
		\min_{\mathclap{\substack{\vec p_1, \dots, \vec p_k,\\\vec p_c,  \vec C, \vec \gamma_p,\\ s, \vec d, \vec e, t, \vec\alpha}}}\quad & t \\
	\mbox{s.t.}\quad
	& \sqrt{s} \big( \sum\limits_{j\in\mathcal{K}}|\vec{h}_{1}^{H}\vec{p}_{j}|^2+1 \big)^{1/2} - {\vec{h}}_{1}^{H}\vec{p}_{c} \le t \label{eq:sitdual2:1}\\
	& \sqrt{s} \big( \sum\limits_{j\in\mathcal{K}}|\vec{h}_{k}^{H}\vec{p}_{j}|^2+1 \big)^{1/2} - d_k \le t,\enskip k > 1 \label{eq:sitdual2:2}\\
	& \sqrt{\gamma_{p,k}} \big( \sum\limits_{j\in\mathcal{K}\setminus k}|\vec{h}_{k}^{H}\vec{p}_{j}|^2+1 \big)^{1/2}\!\!\!\!\! - {\vec{h}}_{k}^{H}\vec{p}_{k},\enskip k\in\mathcal K \label{eq:sitdual2:3}\\
	& (e_k, d_k - t, \alpha_k)\in\tilde{\mathcal C},\enskip k > 1 \label{eq:sitdual2:4} \\
	& \text{\cref{eq:srmax:4,eq:srmax:5,eq:srmax:7,eq:srmax:9,eq:srmaxequiv:4,eq:srmaxequiv:6,eq:sitdual:soc}} \label{eq:sitdual2:5}
	\end{align}
\end{subequations}
with
\begin{equation}
	\tilde{\mathcal C} = \{ e\in\mathds C, d \in \mathds R, \alpha \in\mathds R : d  \le | e |,\enskip \angle e = \alpha \}.
\end{equation}
Note that this is a convex optimization problem for fixed $(\vec\gamma_p, s, \vec\alpha)$. Hence, we design the \cgls{brb} procedure to operate on these variables.

Relating this to the previous section, we can identify the nonconvex variables $\vec x = (\vec\gamma_p, s, \vec\alpha)$, the convex variables $\vec\xi = (\vec p_c, \vec p_1, \dots, \vec p_K, \vec C, \vec\gamma_p, \vec d, \vec e)$, the dual feasible set $\mathcal F$ as
\begin{equation} \label{eq:sitf}
	\begin{aligned}
	&\Big\{ \vec\gamma_p, s, \vec \alpha, \vec p_c, \vec p_1, \dots, \vec p_K, \vec C, \vec\gamma_p, \vec d, \vec e :
		\angle\vec e = \vec\alpha,\\
	&	\quad\vec\alpha\in[0, 2\pi]^{K-1}\!,
	\text{ and \cref{eq:srmax:4,eq:srmax:5,eq:srmax:7,eq:srmax:9,eq:srmaxequiv:4,eq:srmaxequiv:6,eq:sitdual:soc}} \mathrlap{\Big\}.}\enskip
	\end{aligned}
\end{equation}
and the dual objective $\tilde g(\vec x, \vec\xi) = \max_i\{ g_i(\vec x, \vec\xi) \}$ as the function
\begin{equation} \label{eq:sitg}
		\tilde g : (\vec\gamma_p, s, \vec\alpha) \mapsto \min_{\substack{\vec p_1, \dots, \vec p_K,\\ \vec p_c, \vec C, \vec d, \vec e, t}}\  t
		\quad\mathrm{s.\,t.}\quad
		\text{\cref{eq:sitdual2:1,eq:sitdual2:2,eq:sitdual2:3,eq:sitdual2:4,eq:sitdual2:5}}.
\end{equation}

\subsection{Bounding Procedure} \label{sec:sit:bounding}
A bounding function $\beta(\mathcal M)$ that satisfies \cref{eq:consistent} is required. We obtain it by adding suitable box constraints to \cref{eq:sitdual2} and then relaxing it adequately.
First, observe that the objective of \cref{eq:sitdual} is increasing in $(\vec\gamma_p, s)$. Hence, a lower bound on $[\ubar{\vec\gamma}_p, \bar{\vec\gamma}_p] \times [\ubar s, \bar s]$ is obtained by setting $\vec\gamma_p = \ubar{\vec\gamma}_p$ and $s = \ubar s$.
This leaves the nonconvexity in $\tilde{\mathcal C}$. Consistent bounding over this set is achieved by using argument cuts \cite{Lu2017}, i.e., we introduce box constraints on $\vec\alpha$, i.e., $\vec\alpha\in[\ubar{\vec\alpha}, \bar{\vec\alpha}]$, and replace $\tilde{\mathcal C}$ by its convex envelope.
For $\bar\alpha_k - \ubar{\alpha}_k \le \pi$, this envelope is
\begin{subequations}
	\label{eq:crelax}
\begin{gather}
	\sin(\ubar\alpha_k) \Re\{e_k\} - \cos(\ubar\alpha_k) \Im\{e_k\} \le 0 \label{eq:crelax:1} \\
	\sin(\bar\alpha_k) \Re\{e_k\} - \cos(\bar\alpha_k) \Im\{e_k\} \ge 0 \label{eq:crelax:2} \\
	a_k \Re\{e_k\} + b_k \Im\{e_k\} \ge (d_k - t) (a_k^2 + b_k^2) \label{eq:crelax:3}
\end{gather}
\end{subequations}
and $(e_k, d_k) \in \mathds C \times \mathds R$ otherwise \cite[Prop.~1]{Lu2017},
where
$a_k = \tfrac{1}{2} \left( \cos(\ubar\alpha_k) + \cos(\bar\alpha_k) \right)$, and
$b_k = \tfrac{1}{2} \left( \sin(\ubar\alpha_k) + \sin(\bar\alpha_k) \right)$.
Then, the bounding problem for a box $\mathcal M = [\ubar{\vec\gamma}_p, \bar{\vec\gamma}_p] \times [\ubar s, \bar s] \times [\ubar{\vec\alpha}, \bar{\vec\alpha}]$ is
\begin{subequations} \label{eq:sitbndfirst}
	\begin{align}
		\min_{\substack{\vec p_c, \vec p_1, \dots, \vec p_K,\\ \vec C, \vec \gamma_p, s, \vec d, \vec e, t}}\quad & t \\
	\mbox{s.t.}\quad
	& \sqrt{\ubar{s}} \big( \sum\limits_{j\in\mathcal{K}}|\vec{h}_{1}^{H}\vec{p}_{j}|^2+1 \big)^{1/2} - {\vec{h}}_{1}^{H}\vec{p}_{c} \le t \label{eq:sitbndfirst:1}\\
	& \sqrt{\ubar{s}} \big( \sum\limits_{j\in\mathcal{K}}|\vec{h}_{k}^{H}\vec{p}_{j}|^2+1 \big)^{1/2} - d_k \le t,\enskip k > 1 \label{eq:sitbndfirst:2}\\
	& \sqrt{\ubar{\gamma}_{p,k}} \big( \sum\limits_{j\in\mathcal{K}\setminus k}|\vec{h}_{k}^{H}\vec{p}_{j}|^2+1 \big)^{1/2}\!\!\!\!\! - {\vec{h}}_{k}^{H}\vec{p}_{k},\enskip k\in\mathcal K \label{eq:sitbndfirst:3}\\
	& \text{\cref{eq:crelax:1,eq:crelax:2,eq:crelax:3}},\enskip k\in\mathcal I_{\mathcal M} \label{eq:sitbndfirst:5}\\
	& \vec\gamma_p \in [\ubar{\vec\gamma}_p, \bar{\vec\gamma}_p],\quad s\in [\ubar s, \bar s], \label{eq:sitbndfirst:4} \\
	& \text{\cref{eq:srmax:4,eq:srmax:5,eq:srmax:7,eq:srmax:9,eq:srmaxequiv:4,eq:srmaxequiv:6,eq:sitdual:soc}},
	\end{align}
\end{subequations}
where, with a slight abuse of notation,
\begin{equation}
	\mathcal I_{\mathcal M} = \Big\{ k\in\mathcal K : k > 1 \land \max_{\ubar{\alpha}, \bar{\alpha} \in \mathcal M} | \bar{\alpha}_k - \ubar{\alpha}_k | \le \pi \Big\}.
\end{equation}

Define the bounding function $\beta(\mathcal M)$ such that it takes the optimal value of \cref{eq:sitbndfirst} if \cref{eq:sitbndfirst} is feasible and $\infty$ otherwise.
This is a suitable bounding function to solve \cref{eq:sitdual} with a \cgls{bb} procedure over the nonconvex variables $\vec{\gamma}_p, s, \vec\alpha$.
\begin{lemma} \label{lem:consist}
	The bounding function $\beta(\mathcal M)$ computed from \cref{eq:sitbndfirst} is consistent with respect to \cref{eq:sitdual2}, i.e., it satisfies \cref{eq:consistent} with $g_i(\vec x, \vec \xi)$ and $\mathcal F$ as identified in \cref{eq:sitf,eq:sitg}.
\end{lemma}
\begin{IEEEproof}
	We need to show that $\beta(\mathcal M)$ asymptotically approaches the optimal value of \cref{eq:sitbndfirst} on $\mathcal M$ as $\mathcal M$ shrinks to a singleton, i.e., $\mathcal M \rightarrow \{ \vec z^* \}$ with $\vec z^* = (\vec\gamma_p^*, s, \vec\alpha^*)$. Observe that these problems only differ in the constraints \cref{eq:sitdual2:1,eq:sitdual2:2,eq:sitdual2:3,eq:sitdual2:4} and \cref{eq:sitbndfirst:1,eq:sitbndfirst:2,eq:sitbndfirst:3,eq:sitbndfirst:5}. Asymptotically, \cref{eq:sitdual2:1,eq:sitdual2:2,eq:sitdual2:3} and \cref{eq:sitbndfirst:1,eq:sitbndfirst:2,eq:sitbndfirst:3} are equivalent since $\ubar{\vec\gamma}_p, \vec\gamma_p \rightarrow \vec\gamma_p^*$ and $\ubar{s}, s \rightarrow s^*$.

	For the remaining constraints, note that $\mathcal I_{\mathcal M} = \{ k\in\mathcal K: k > 1\}$ as $\ubar{\vec\alpha}, \bar{\vec\alpha} \rightarrow \vec\alpha^*$. Further, \cref{eq:crelax:1,eq:crelax:2} asymptotically evaluate to
	\begin{equation} \label{eq:lem:sitconv:1}
		\sin(\alpha_k^*) \Re\{e_k\} = \cos(\alpha_k^*) \Im\{e_k\}
	\end{equation}
	and \cref{eq:crelax:3} to
	\begin{equation} \label{eq:lem:sitconv:2}
		\cos(\alpha_k^*) \Re\{e_k\} + \sin(\alpha_k^*) \Im\{e_k\} \ge d_k - t.
	\end{equation}
	Recall that constraint \cref{eq:sitdual2:4} is $d_k - t \le | e_k |$ and $\angle e_k = \alpha_k^*$. The second equation is equivalent to
	\begin{equation}
		| e_k | = \frac{\Re\{ e_k \}}{\cos(\alpha_k^*)} = \frac{\Im\{ e_k \}}{\sin(\alpha_k^*)}
	\end{equation}
	and, hence, asymptotically the same as \cref{eq:lem:sitconv:1}.
	Finally, with $\Re\{e_k\} = |e_k| \cos(\alpha_k^*)$ and $\Im\{e_k\} = |e_k| \cos(\alpha_k^*)$, the \cgls{lhs} of \cref{eq:lem:sitconv:2} is
	\begin{equation}
		\cos^2(\alpha_k^*) | e_k | + \sin^2(\alpha_k^*) | e_k | = | e_k |.
	\end{equation}
	This establishes the lemma.
\end{IEEEproof}

Observe that problem \cref{eq:sitbndfirst} depends on $\vec\gamma_p$ and $s$ only through constraints \cref{eq:srmaxequiv:4,eq:srmax:9,eq:sitdual:soc,eq:sitbndfirst:4}. These are (convex) exponential cone constraints that can be transformed into affine functions of $(\vec\gamma_p, s)$ by substituting $s' = \log(1 + s)$ and $\gamma_{p,k}' = \log( 1 + \gamma_{p,k})$. This leads to an equivalent optimization problem with considerably reduced computational complexity. In particular, we can solve the following \cgls{socp}
\begin{subequations} \label{eq:sitbnd}
	\begin{align}
		\min_{\substack{\vec p_1, \dots, \vec p_K, \vec p_c,\\ \vec C, \vec \gamma_p', s', \vec d, \vec e, t}} & t \\
	\mbox{s.t.}\quad
	& C_k \geq \max \left\{ 0,\ R_k^{th} - \gamma_{p,k}' \right\},\enskip  k\in\mathcal{K} \\
	& \sum\nolimits_{k\in \mathcal{K}}C_{k}\leq s' \\
	& \sum\nolimits_{k\in\mathcal{K}} u_k \left(C_k+\gamma_{p,k}' \right) \notag\\
	& \qquad\ge \delta \Big( \mu \Big( \Vert \vec p_c \Vert^2 + \sum\limits_{k\in\mathcal K} \Vert \vec p_k \Vert^2 \Big) + P_c \Big) \\
	& \gamma'_{p,k} \!\in\! [\log(1+\ubar{\gamma}_{p,k}), \log(1+\bar{\gamma}_{p,k})], k\in\mathcal K \\
	& s'\in [\log(1+\ubar s), \log(1+\bar s)], \\
	& \text{\cref{eq:srmax:4,eq:srmax:5,eq:srmax:7,eq:srmaxequiv:6,eq:sitbndfirst:1,eq:sitbndfirst:2,eq:sitbndfirst:3,eq:sitbndfirst:5}}
	\end{align}
\end{subequations}
instead of \cref{eq:sitbndfirst} to compute the bounding function $\beta(\mathcal M)$.

\subsection{Feasible Point} \label{sec:sit:feas}
In every iteration, a dual feasible point $\vec x^k$ is required that satisfies $\vec x^k \in\proj_{\vec x} \mathcal F \cap \mathcal M_k$ whenever $\proj_{\vec x} \mathcal F\cap\mathcal M_k \neq \emptyset$ (or, equivalently, $\beta(\mathcal M_k) < \infty$). If this point satisfies \cref{eq:sit:primalfeas}, it is nonisolated primal feasible
according to \cref{lem:conv} and can be used to update $\delta$ in \cref{alg:sit}.
Such a point $(\vec\gamma_p^k, s^k, \vec\alpha^k)$ can be obtained from the
optimal solution $(\vec\gamma_p^\star, s^\star, \vec e^\star, \dots)$ of the bounding problem \cref{eq:sitbnd} as
\begin{subequations}\label{eq:xk}
\begin{align} \label{eq:xk:1}
	\gamma_{p,i}^k &= 2^{\gamma_{p,i}^\star}-1, i\in\mathcal K,
	& s^k = 2^{s^\star}-1
\end{align}
and $\vec\alpha^k\in\proj_{\vec\alpha} \mathcal M_k = [\ubar{\vec\alpha}^k, \bar{\vec\alpha}^k]$. At first glance, a sensible choice for $\vec\alpha$ seems to be $\alpha_i^k = \angle e_i^\star$. However, preliminary numerical experiments show that this point leads to very slow convergence. Much better results are obtained by using the corner point of $\proj_{\vec\alpha}\mathcal M^k$ closest to $\angle\vec e^\star$, i.e.,
\begin{equation} \label{eq:xk:2}
	\alpha^k_i = \argmin_{\alpha\in\{\ubar{\alpha}_i^k, \bar{\alpha}_i^k\}} |\alpha - \angle e_i^\star|.
\end{equation}
\end{subequations}
It is easily verified that this point satisfies
$(\vec\gamma_p^k, s^k, \vec\alpha^k)\in\proj_{(\vec\gamma_p, s, \vec\alpha)} \mathcal F\cap\mathcal M_k$. If further
$\tilde g(\vec\gamma_p^k, s^k, \vec\alpha^k) \le 0$, it is 
primal feasible and the solution of \cref{eq:sitg} achieves a primal objective value greater than or equal to $\delta$.
\begin{lemma} \label{lem:sitfeas}
	Let $\vec z$ be a solution of \cref{eq:sitbnd} for some $\delta$.
	Obtain $\vec x^k = (\vec \gamma_p^k, s^k, \vec\alpha^k)$ from $\vec z$ as in \cref{eq:xk}.
	Compute $\tilde g(\vec\gamma_p^k, s^k, \vec\alpha^k)$ as in \cref{eq:sitg} and
	let $(t^\star, \vec y^\star) = (t^\star, \vec p_1^\star, \dots, \vec p_K^\star, \vec p_c^\star, \vec C^\star, \vec d^\star, \vec e^\star)$ be a solution of the accompanying optimization problem.
	If $t^\star \le 0$, $(\vec x^k, \vec y^\star)$ is a primal feasible point with primal objective value greater than or equal to $\delta$.
	Then, $(\vec y^\star, \vec \gamma_p^\star, s^\star)$ with $\gamma_{p,k}^\star = \frac{|{\vec{h}}_{k}^{H}\vec{p}_{k}^\star|^2}{\sum\limits_{j\in\mathcal{K}\setminus k}|\vec{h}_{k}^{H}\vec{p}_{j}^\star|^2+1}$ for all $k\in\mathcal{K}$ and $s^\star = \min_{k\in\mathcal K} \frac{|{\vec{h}}_{k}^{H}\vec{p}_{c}^\star|^2}{\sum\limits_{j\in\mathcal{K}}|\vec{h}_{k}^{H}\vec{p}_{j}^\star|^2+1}$ is a feasible point of \cref{eq:srmax} and achieves a primal objective value greater than or equal to that of $(\vec x^k, \vec y^\star)$.
	The primal objective value can be further improved (while preserving primal feasibility) by updating $\vec C^\star$ to a solution of
	\begin{subequations} \label{eq:bestC}
		\begin{align}
			\max_{\vec C} \quad & \sum_{k\in\mathcal K} u_k C_k \\
		\mbox{s.t.}\quad
	&  \frac{\sum_{k\in\mathcal{K}} u_k \left( C_k + \log(1+\gamma_{p,k}^\star) \right)}{\mu \left( \Vert \vec p_c^\star \Vert^2 + \sum_{k\in\mathcal K} \Vert \vec p_k^\star \Vert^2 \right) + P_c} \ge \delta \label{eq:bestC:1} \\
		& \sum_{k\in \mathcal{K}}C_{k}\leq \log(1+s^\star) \label{eq:bestC:2} \\
		& C_k \geq \max \left\{ 0,\ R_k^{th} - \log(1+\gamma_{p,k}^\star) \right\}, \forall k\in\mathcal{K}. \label{eq:bestC:3}
		\end{align}
	\end{subequations}
\end{lemma}

\begin{IEEEproof}
	Observe that $\vec\gamma_p^k$ and $s^k$ are such that the set defined by \cref{eq:srmaxequiv:4,eq:srmax:9,eq:sitdual:soc,eq:srmax:4,eq:srmax:5,eq:srmax:7,eq:srmaxequiv:6} is nonempty. Further, if $t \le 0$, every point satisfying \cref{eq:sitdual2:1,eq:sitdual2:2,eq:sitdual2:3} also meets \cref{eq:srmax:1,eq:srmax:2,eq:srmax:3}. Since $\tilde{\mathcal C} \subseteq \mathcal C$, \cref{eq:sitdual2:4} implies that $(e_k, d_k)\in\mathcal C$ if $t \le 0$. Hence, $(\vec y^\star, \vec \gamma_p^k, s^k)$ is a feasible solution of \cref{eq:srmax} if $t^\star \le 0$.

	From the proof of \cref{prop:srmax}, we know that every point that satisfies \cref{eq:srmax:1,eq:srmax:2,eq:srmax:3,eq:srmax:4,eq:srmax:5,eq:srmax:7,eq:srmax:8} also satisfies \cref{eq:srmaxinter:1,eq:srmaxinter:2}.
	This implies $\vec\gamma_p^k \le \vec\gamma_p^\star$, $s^k \le s^\star$ and, hence, $(\vec\gamma_p^\star, \vec s^\star)$ satisfies \cref{eq:srmax:1,eq:srmax:2,eq:srmax:3,eq:srmax:4,eq:srmax:5,eq:srmax:7,eq:srmax:8,eq:srmax:9,eq:srmaxequiv:4}. Thus, $(\vec y^\star, \vec \gamma_p^\star, s^\star)$ is a feasible point of \cref{eq:srmax}.

	Clearly, $(\vec x^k, \vec y^\star)$ satisfies \cref{eq:sitdual:soc}. Hence,
	\begin{multline}
	\delta
		\le \frac{\sum_{i\in\mathcal K} u_i ( C_i^\star + \log( 1 + \gamma_{p,i}^k) )}{{\mu \left( \Vert \vec p_c^\star \Vert^2 + \sum_{i\in\mathcal K} \Vert \vec p_i^\star \Vert^2 \right) + P_c}} \\
		\le \frac{\sum_{i\in\mathcal K} u_i ( C_i^\star + \log( 1 + \gamma_{p,i}^\star) )}{{\mu \left( \Vert \vec p_c^\star \Vert^2 + \sum_{i\in\mathcal K} \Vert \vec p_i^\star \Vert^2 \right) + P_c}}. \label{lem:sitfeas:proof:1}
	\end{multline}
	Clearly, any $\vec C$ that is feasible in \cref{eq:bestC} is also feasible in \cref{eq:srmax} and maximizes \cref{lem:sitfeas:proof:1} in $\vec C$ (with all other variables fixed).
\end{IEEEproof}

\subsection{Initial Box} \label{sec:initbox}
The \cgls{bb} procedure requires an initial box $\mathcal M_0 = [\ubar{\vec\gamma}_p^0, \bar{\vec\gamma}_p^0] \times [\ubar s^0, \bar s^0] \times [\ubar{\vec\alpha}^0, \bar{\vec\alpha}^0]$ that contains the nonconvex dimensions of the dual feasible set $\proj_{(\vec\gamma_p, s, \vec\alpha)} \mathcal F$. As $\vec\alpha$ is already constrained by box constraints, we have $[\ubar{\vec\alpha}^0, \bar{\vec\alpha}^0] = [0, 2\pi]^{K-1}$. For $\bar{\vec\gamma}_p^0$, observe that $\bar\gamma_{p,k}^0 \ge \max_{\vec\gamma_p, s, \vec\alpha\in\mathcal F} \gamma_{p,k}$ but also
\begin{equation}
	\bar\gamma_{p,k}^0 \ge \max_{\substack{\vec p_c, \vec p_1, \dots, \vec p_K, \vec C, \vec \gamma_c, \vec \gamma_p}} \gamma_{p,k} \quad\st\quad \text{\cref{eq:srmaxequiv:2,eq:srmaxequiv:3,eq:srmaxequiv:4,eq:srmaxequiv:6}}.
	\label{eq:intialboxmax}
\end{equation}
This nonconvex optimization problem can be relaxed to
\begin{equation}
	\max_{\vec p_k}\enskip |{\vec{h}}_{k}^{H}\vec{p}_{k}|^2 \quad\st\quad \Vert \vec p_k \Vert^2 \le P. \label{eq:initialboxmax2}
\end{equation}
The solution to \cref{eq:initialboxmax2} is $\vec p_k^\star = \sqrt{P} \frac{\vec h_k}{\Vert\vec h_k\Vert}$ \cite[\S 5.3.2]{Tse2005} and, hence, $\bar\gamma_{p,k}^0 = P \Vert \vec h_k \Vert^2$.
Likewise, the upper bound $\bar s^0$ for $s$ needs to satisfy
\begin{equation}\label{eq:sitsboxinter}
	\bar s^0 \ge 
	\max_{\vec p_c}\enskip \min_{k\in\mathcal K}\ |{\vec{h}}_{k}^{H}\vec{p}_{c}|^2
\quad\st\quad
\Vert \vec p_c \Vert^2 \le P.
\end{equation}
This is an NP-hard optimization problem as discussed in \cref{sec:sitfund}. Exchanging maximum and minimum leads to the relaxed problem
\begin{equation}
	\bar s^0 = \min_{k\in\mathcal K}\ \max_{\Vert \vec p_c \Vert^2 \le P}\ |{\vec{h}}_{k}^{H}\vec{p}_{c}|^2 = \min_{k\in\mathcal K} P \Vert \vec h_k \Vert^2.
\end{equation}

An obvious lower bound on $\vec\gamma_p$ and $s$ is 0. For $\vec\gamma_p$, we can exploit the \cgls{qos} constraints to obtain a possibly tighter initial box. Similar to the upper bound, $\ubar{\gamma}_{p,k}$ needs to be less than or equal to the minimum of $\gamma_{p,k}$ over \cref{eq:srmaxequiv:2,eq:srmaxequiv:3,eq:srmaxequiv:4,eq:srmaxequiv:6}. The optimal solution to this problem either meets \cref{eq:srmaxequiv:4} with equality or is zero. In the first case, this is equivalent to
\begin{subequations} \label{eq:gammamininter}
	\begin{align}
		\min_{\vec p_c, \vec p_1, \dots, \vec p_K, \vec C}\quad & 2^{R_k^{th} - C_k} - 1 \\
	\mbox{s.t.}\quad
	& \sum_{k'\in \mathcal{K}}C_{k'}\leq \log\bigg(1+ \min_{k\in\mathcal K}\frac{|{\vec{h}}_{k}^{H}\vec{p}_{c}|^2}{\sum\limits_{j\in\mathcal{K}}|\vec{h}_{k}^{H}\vec{p}_{j}|^2+1}\bigg) \\
	& \vec C\geq \vec{0},\quad
	\Vert \vec p_c \Vert^2 + \sum_{k\in\mathcal K} \Vert \vec p_k \Vert^2 \le P.
	\end{align}
\end{subequations}
Clearly, the optimal solution to this problem is equivalent to the solution of
\begin{subequations}
	\begin{align}
	\max_{\vec p_c, \vec p_1, \dots, \vec p_K, \vec C}\quad & \log\bigg(1+ \min_{k\in\mathcal K}\frac{|{\vec{h}}_{k}^{H}\vec{p}_{c}|^2}{\sum\limits_{j\in\mathcal{K}}|\vec{h}_{k}^{H}\vec{p}_{j}|^2+1}\bigg) \\
	\mbox{s.t.}\quad
	& \Vert \vec p_c \Vert^2 + \sum_{k\in\mathcal K} \Vert \vec p_k \Vert^2 \le P.
	\end{align}
\end{subequations}
The optimal choice for $\vec p_1, \dots, \vec p_K$ is $\vec 0$. Optimizing over $\vec p_c$ is equivalent to \cref{eq:sitsboxinter}. Thus, an upper bound to the optimal $C_k$ is $\log(1+\bar{s})$ and a lower bound on the optimal value of \cref{eq:gammamininter} is $2^{R_k^{th} - \log(1+\bar{s})} - 1$. Hence,
\begin{equation}
\ubar{\gamma}_{p,k} = \max\bigg\{ 0, \frac{2^{R_k^{th}}}{ 1 + \bar{s}} - 1 \bigg\}.
\end{equation}

\subsection{Reduction Procedure} \label{sec:sit:red}
The convergence criterion \cref{eq:consistent} implies that the quality of the bound $\beta(\mathcal M)$ improves as the diameter of $\mathcal M$ shrinks. Since tighter bounds lead to faster convergence, it is beneficial to reduce the size of $\mathcal M$ prior to bounding if possible at reasonable computational cost. It is important that such a reduced box $\mathcal M' \subseteq \mathcal M$ still contains all solution candidates, i.e., $\mathcal M \cap \tilde{\mathcal F} = \mathcal M' \cap \tilde{\mathcal F}$ or, equivalently,
$(\mathcal M\setminus\mathcal M') \cap \tilde{\mathcal F} = \emptyset$, where $\tilde{\mathcal F} = \proj_{(\vec\gamma_p, s, \vec\alpha)} \mathcal F$.

A suitable reduction is derived in the lemma below. Preliminary numerical experiments have shown that this procedure is essential to ensure convergence within reasonable time.

\begin{lemma} \label{lem:reduction}
	Let $\mathcal M = [\ubar{\vec\gamma}_p, \bar{\vec\gamma}_p]\times[\ubar s, \bar s]\times[\ubar{\vec\alpha}, \bar{\vec\alpha}]$, $\mathcal M' =  [\ubar{\vec\gamma}_p', \bar{\vec\gamma}_p']\times[\ubar s', \bar s']\times[\ubar{\vec\alpha}, \bar{\vec\alpha}]$, and $\tilde{\mathcal F} = \proj_{(\vec\gamma_p, s, \vec\alpha)} \mathcal F$. Then, $\mathcal M \cap \tilde{\mathcal F} = \mathcal M' \cap \tilde{\mathcal F}$ if
	\begin{align*}
		\ubar{\gamma}_{p,k}' &= \max\{ \ubar{\gamma}_{p,k}, \ubar{\gamma}_{p,k}'' \} \\
		\bar{\gamma}_{p,k}' &= \min\Big\{ \bar{\gamma}_{p,k},\ \ubar{\gamma}_{k,p}' + \frac{\Vert \vec h_k\Vert^2}{\delta\mu} ( U - \delta W') \Big\} \\
		\ubar s' &= \max\left\{ \ubar s,\ 2^{\max\big\{\frac{W \delta - U}{\max_{k\in\mathcal K} \{u_k\}},\ V\big\}} (1+ \bar s) - 1 \right\} \\
		\bar{s}' &= \min\Big\{ \bar s,\ \ubar{s}' + \frac{\min_k \Vert \vec h_k\Vert^2}{\delta\mu} ( U - \delta W') \Big\}
	\end{align*}
	with
	\begin{equation*}
		\ubar{\gamma}_{p,k}'' = 
		\begin{cases}
			2^{\max\{ \frac{W \delta - U}{u_k},\ V\}} (1+ \bar\gamma_{p,k}) - 1, & k\in\mathcal I \\
			\max\{ \ 2^{\frac{W \delta - U}{u_k}} (1+ \bar\gamma_{p,\kappa}), 2^{V+R_k^{th}}  \} - 1, & k\notin\mathcal I
		\end{cases}
	\end{equation*}
	and
	\begin{align*}
		\mathcal I &= \{ k \in\mathcal K : R_k^{th} - \log(1+ \bar\gamma_{p,k}) > 0 \} \\
		U &= \max_{k\in\mathcal K} \{ u_k \} \log(1+ \bar s) + \sum_{k\in\mathcal{K}} u_k \log(1+ \bar\gamma_{p,k}) \\
		V &= \sum_{k\in \mathcal{I}} \left( R_k^{th} - \log(1+ \bar\gamma_{p,k}) \right) -  \log(1+ \bar s) \\
		W &= \mu \Big( \ubar s \max_k \Vert \vec h_k \Vert^{-2} + \sum\nolimits_{k\in\mathcal K} \ubar{\gamma}_{p,k}\Vert \vec h_{k} \Vert^{-2} \Big) + P_c \\
		W' &= \mu \Big( \ubar s' \max_k \Vert \vec h_k \Vert^{-2} + \sum\nolimits_{k\in\mathcal K} \ubar{\gamma}_{p,k}'\Vert \vec h_{k} \Vert^{-2} \Big) + P_c.
	\end{align*}
\end{lemma}

\begin{IEEEproof}
	Due to monotonicity, a necessary condition for $\mathcal M \cap \tilde{\mathcal F} \neq \emptyset$ is that
	\cref{eq:srmax:9,eq:srmaxequiv:4,eq:sitdual:soc}
	hold when evaluated at $(\bar{\vec\gamma}_p, \bar{s}, \bar{\vec\alpha})$.
	For \cref{eq:sitdual:soc}, this implies
	\begin{subequations} \label{eq:redmin1}
	\begin{align}
		\MoveEqLeft \max_{k\in\mathcal K}  \sum_{k\in\mathcal{K}}  u_k C_k + \sum_{k\in\mathcal{K}} u_k \log(1+ \bar\gamma_{p,k}) \\
		&\ge  \delta \bigg( \mu \bigg( \Vert \vec p_c \Vert^2 + \sum_{k\in\mathcal K} \Vert \vec p_k \Vert^2 \bigg) + P_c \bigg)  \\
		&\ge  \delta \!\min_{\vec p_c, \vec p_1, \dots, \vec p_K}\!\bigg\{ \mu \bigg( \Vert \vec p_c \Vert^2 + \sum_{k\in\mathcal K} \Vert \vec p_k \Vert^2 \bigg) + P_c \bigg\} \label{eq:redmin} \\
		&\ge  \delta \bigg( \mu \bigg( \min \Vert \vec p_c \Vert^2 + \sum_{k\in\mathcal K} \min \Vert \vec p_k \Vert^2 \bigg) + P_c \bigg) \label{eq:redmin2}
	\end{align}
	\end{subequations}
	where the minimum in \cref{eq:redmin,eq:redmin2} is such that $\vec\gamma_p\in\mathcal M$, i.e., for all $\kappa = 1, \dots, K$,
	\begin{equation*}
		\min_{\vec p_c, \dots, \vec p_K} \Vert \vec p_\kappa \Vert^2 \quad\st\quad \forall k : \ubar\gamma_{p,k} \le \frac{|{\vec{h}}_{k}^{H}\vec{p}_{k}|^2}{\sum\limits_{j\in\mathcal{K}\setminus k}|\vec{h}_{k}^{H}\vec{p}_{j}|^2+1} \le \bar\gamma_{p,k}.
	\end{equation*}
	This can be relaxed to
	\begin{equation} \label{eq:redmin3}
		\min_{\vec p_c, \dots, \vec p_K} \Vert \vec p_k \Vert^2 \quad\st\quad \ubar\gamma_{p,k} \le |{\vec{h}}_{k}^{H}\vec{p}_{k}|^2.
	\end{equation}
	After transforming \cref{eq:redmin3} into a \cgls{socp}, an optimal solution can be readily obtained from the \cgls{kkt} conditions as $\vec p_{\kappa}^\star = \sqrt{\ubar{\gamma}_{p,\kappa}} \frac{\vec h_\kappa}{\Vert \vec h_\kappa \Vert^2}$ with optimal value $\frac{\ubar{\gamma}_{p,\kappa}}{\Vert \vec h_{\kappa} \Vert^2}$. Likewise, a lower bound for $\Vert\vec p_c\Vert^2$ is obtained as $\frac{\ubar s}{\min_k \Vert \vec h_k \Vert^2}$. Combining this with \cref{eq:redmin1} and \cref{eq:srmax:9}, we get the necessary condition $U \ge W \delta$.

	Let $\mathcal M'' = [\ubar{\vec\gamma}_p', \bar{\vec\gamma}_p]\times[\ubar s', \bar s]\times[\ubar{\vec\alpha}, \bar{\vec\alpha}]$. It follows from $U \ge W \delta$, that
	every dual feasible $\gamma_{p,\kappa}$ in $\mathcal M$ satisfies
	\begin{equation} \label{eq:redW}
		W \delta
		\le U - u_\kappa \log(1+\bar\gamma_{p,\kappa})  + u_\kappa \log(1+ \gamma_{p,\kappa}).
	\end{equation}
	This is equivalent to
	\begin{equation} \label{eq:red1}
		\gamma_{p,\kappa} \ge 2^{\frac{W \delta - U}{u_\kappa}} (1+ \bar\gamma_{p,\kappa}) - 1.
	\end{equation}
	Hence, every $\gamma_{p,\kappa}\in\proj_{\gamma_{p,\kappa}}(\mathcal M \cap \tilde{\mathcal F})$ needs to satisfy \cref{eq:red1}.
	From the initial remark, we further observe that \cref{eq:srmax:9,eq:srmaxequiv:4} can only hold if $V \le 0$.
	Thus, every $\gamma_{p,\kappa}\in\mathcal M\cap\tilde{\mathcal F}$ with $\kappa\in\mathcal I$ satisfies
	\begin{flalign}
		 && \log(1+ \gamma_{p,\kappa}) &\ge V + \log(1+ \bar\gamma_{p,\kappa}) && \\
		 \Leftrightarrow&& \gamma_{p,\kappa} &\ge 2^V (1+ \bar\gamma_{p,\kappa}) - 1,
	\end{flalign}
	and every $\gamma_{p,\kappa}\in\proj_{\gamma_{p,\kappa}}(\mathcal M \cap \tilde{\mathcal F})$ with $\kappa\notin\mathcal I$ satisfies
	\begin{flalign}
		&&  \log(1+ \gamma_{p,\kappa}) &\ge V + R_\kappa^{th} && \\
		\Leftrightarrow && \gamma_{p,\kappa} &\ge 2^{V + R_\kappa^{th}} - 1.
	\end{flalign}
	This establishes $\ubar{\vec\gamma}_p'$. The lower bound $\ubar{s}'$ for $s$ is obtained analogously.

	Further, every $\gamma_{p,\kappa}\in\proj_{\gamma_{p,\kappa}}(\mathcal M'' \cap \mathcal F)$ satisfies
	\begin{align}
		&\delta \Big( W' + \frac{\mu}{\Vert \vec h_\kappa\Vert^2} \left( \gamma_{\kappa,p} - \ubar{\gamma}_{\kappa,p}' \right) \Big) \le U \\
		\Leftrightarrow\quad &\gamma_{k,p} \le \ubar{\gamma}_{\kappa,p}' + \frac{\Vert \vec h_\kappa\Vert^2}{\delta\mu} ( U - \delta W').
	\end{align}
	Similarly, every $s\in\proj_{s}(\mathcal M'' \cap \mathcal F)$ satisfies
	\begin{align}
		&\delta \Big( W' + \frac{\mu}{\min_k \Vert \vec h_\kappa\Vert^2} \left( s - \ubar{s}' \right) \Big) \le U \\
		\Leftrightarrow\quad &s \le s' + \frac{\min_k \Vert \vec h_\kappa\Vert^2}{\delta\mu} ( U - \delta W').
	\end{align}
	Hence, the upper bounds on $\vec\gamma_p$ and $\vec s$ can be reduced to $\bar{\gamma}_p'$ and $\bar{s}'$, respectively.
\end{IEEEproof}

\begin{corollary} \label{corr:feas}
	Let $\mathcal M$, $\tilde{\mathcal F}$, $U$, $V$, and $W$ be as in \cref{lem:reduction}. Then, $\mathcal M$ is infeasible, i.e., $\mathcal M \cap \tilde{\mathcal F} = \emptyset$, if $V > 0$ or $U < W \delta$.
\end{corollary}
\begin{IEEEproof}
From the proof of \cref{lem:reduction}, we know that every $\mathcal M \cap \tilde{\mathcal F} \neq \emptyset$ satisfies $V \le 0$ and $U \ge W \delta$.
\end{IEEEproof}

\subsection{Algorithm and Convergence}
The complete algorithm is stated in \cref{alg:sitbb}. It is essentially a \cgls{brb} procedure \cite{Tuy2016,diss} that solves the \cgls{sit} dual of \cref{eq:srmax} and updates the constant $\delta$ whenever a primal feasible point is encountered.

\begin{algorithm}
\renewcommand{\crefrangeconjunction}{--}
	\caption{\cGls{sit} Algorithm for \cref{eq:srmax}}\label{alg:sitbb}
	\small
	\centering
	\begin{minipage}{\linewidth-1em}
	\begin{enumerate}[label=\textbf{Step \arabic*},ref=Step~\arabic*,start=0,leftmargin=*]
		\item\label{alg:sitbb:init} {\bfseries (Initialization)} Set $\varepsilon, \eta > 0$. 
			Let $k=1$ and $\mathscr R_0 = \{ \mathcal M_0 \}$ with $\mathcal M_0$ as in \cref{sec:initbox}.
			If an initial feasible solution $\vec y^0 = (\vec p_c^0, \dots, \vec p_K^0)$ is available, set $\delta_0 = \eta + v\eqref{eq:srmaxequiv}|_{\vec y_0}$ and initialize $\bar{\vec x}^0 = (\vec\gamma_p^0, s^0, \vec\alpha^0)$ with \cref{eq: SINR cp}, $s^0 = \min_{i\in\mathcal K} \gamma_{c,i}^0$, and $\alpha^0_i = \angle \vec h_{i}^H \vec p_c^0$, $i>1$.
			Otherwise, do not set $\bar{\vec x}^0$ and choose $\delta_0 = 0$.
		\item\label{alg:sitbb:branch} {\bfseries (Branching)} Let $$\mathcal M_k = [\vec r^k, \vec s^k] = \argmin\{\beta(\mathcal M) \,|\, \mathcal M\in\mathscr R_{k-1} \}.$$ Bisect $\mathcal M_k$ via $(\vec v^k, j_k)$ where $j_k \in \argmax_j s^k_j - r^k_j$ and $\vec v^k = \frac{1}{2} (\vec s^k + \vec r^k)$ as in \cref{eq:partition} and set $\mathscr P_{k} = \{\mathcal M^k_-, \mathcal M^k_+\}$.
		\item\label{alg:sitbb:reduction} {\bfseries (Reduction)} Replace each box $\mathcal M\in\mathscr P_k$ with $\mathcal M'$ as in \cref{lem:reduction}.
		\item\label{alg:sitbb:bounding} {\bfseries (Bounding)} For each reduced box $\mathcal M\in\mathscr P_k$, perform a preliminary feasibility check with \cref{corr:feas} and, if necessary, solve \cref{eq:sitbnd}. If infeasible, set $\beta(\mathcal M) = \infty$. Otherwise, set $\beta(\mathcal M)$ to the optimal value of \cref{eq:sitbnd} and obtain a dual feasible point $\vec x(\mathcal M)$ as in \cref{eq:xk}.
		\item\label{alg:sitbb:feaspoint} {\bfseries (Feasible Point)} For each $\mathcal M\in\mathscr P_k$, if $\beta(\mathcal M) \le 0$ compute $\tilde g(\vec x(\mathcal M))$ as in \cref{eq:sitg}. If $\tilde g(\vec x(\mathcal M)) \le 0$, $\vec x(\mathcal M)$ is primal feasible. Recover $\vec x'(\mathcal M)$ from the solution of \cref{eq:sitg} with $\vec\gamma_p'$, $s'$ as in \cref{lem:sitfeas} and $\alpha_i' = \angle e_i^*$, $i > 1$, with $\vec e^*$ as in \cref{lem:sitfeas}. Compute the primal objective value $f(\mathcal M) = \sum_{j\in\mathcal K} u_k (\tilde{C}_j^\star + \log(1+\gamma_{p,j}'))$, where $\tilde{\vec C}^*$  is the optimal solution of \cref{eq:bestC}. If $\beta(\mathcal M) > 0$ or $\tilde g(\vec x(\mathcal M)) > 0$, set $f(\mathcal M) = -\infty$.
		\item\label{alg:sitbb:incumbent} {\bfseries (Incumbent)} Let $\mathcal M' \in \argmax\{ f(\mathcal M) : \mathcal M\in\mathscr P_k \}$. If $f(\mathcal M') > \delta_{k-1} - \eta$, set $\bar{\vec x}^k = \vec x'(\mathcal M')$ and $\delta_k = f(\mathcal M') + \eta$. Otherwise, set $\bar{\vec x}^k = \bar{\vec x}^{k-1}$ and $\delta_k = \delta_{k-1}$.
		\item\label{alg:sitbb:prune} {\bfseries (Pruning)} Delete every $\mathcal M\in\mathscr P_k$ with $\beta(\mathcal M) > -\varepsilon$. Let $\mathscr P_k'$ be the collection of remaining sets and set $\mathscr R_k = \mathscr P_k'\cup(\mathscr R_{k-1}\setminus\{\mathcal M_k\})$.
		\item\label{alg:sitbb:terminate} {\bfseries (Termination)} Terminate if $\mathscr R = \emptyset$: If $\bar{\vec x}^k$ is not set, then \cref{eq:srmax} is $\varepsilon$-essential infeasible; else $\bar{\vec x}^k$ is an essential $(\varepsilon, \eta)$-optimal solution of \cref{eq:srmax}. Otherwise, update $k\gets k+1$ and return to \ref{alg:sitbb:branch}.
	\end{enumerate}
	\end{minipage}
\end{algorithm}

The algorithm is initialized in \ref{alg:sitbb:init}. The initial box $\mathcal M_0$ is computed as discussed in \cref{sec:initbox}. The set $\mathscr R_k$ holds the current partition of the feasible set, $\delta_k$ is the \cgls{cbv} adjusted by the tolerance $\eta$, and $\bar{\vec x}^k$ is the \cgls{cbs}. If a primal feasible solution $\vec y^0$ is known, it can be used to hot start the algorithm where the variables $\vec\gamma_p^0$ and $s$ are initialized from $\vec y^0$ as in \cref{lem:sitfeas}. Observe that $\vec y^0$ needs to satisfy \cref{eq:srmax:4,eq:srmax:5}. In \ref{alg:sitbb:branch}, the box most likely to contain a good feasible solution is selected as $\mathcal M_k$ and bisected. The new boxes are stored in $\mathscr P_k$ and reduced according to \cref{lem:reduction} in \ref{alg:sitbb:reduction}. The reduced boxes replace the original boxes in $\mathscr P_k$. In \ref{alg:sitbb:bounding}, bounds for each box in $\mathscr P_k$ are computed, infeasibility is detected, and dual feasible points are obtained from the bounding problem (cf.~\cref{sec:sit:bounding,sec:sit:feas}). For each of these dual feasible points, primal feasibility is checked in \ref{alg:sitbb:feaspoint}. If true, a primal feasible point is recovered as established in \cref{lem:sitfeas} and the corresponding primal objective value is computed. Should any of these feasible points achieve a higher objective value than the \cgls{cbs}, the \cgls{cbs} and $\delta_k$ are updated in \ref{alg:sitbb:incumbent}. Boxes that cannot contain primal $\varepsilon$-essential feasible solutions are pruned in \ref{alg:sitbb:prune}. If the partition $\mathscr R_k$ contains undecided boxes, the algorithm is continued in \ref{alg:sitbb:terminate}.

Convergence of the algorithm follows from the previous discussion and is formally established next.
\begin{theorem} \label{thm:sitbb}
	\Cref{alg:sitbb} converges in finitely many steps to the $(\varepsilon, \eta)$-optimal solution of \cref{eq:srmax} or establishes that no such solution exists.
\end{theorem}
\begin{IEEEproof}
	\renewcommand{\crefrangeconjunction}{--}
	\Cref{lem:reduction} ensures that no feasible solution candidates with objective values greater than $\delta_k$ are lost in \ref{alg:sitbb:reduction}.
	The bisection in \ref{alg:sitbb:branch} in exhaustive \cite[Cor.~6.2]{Tuy2016}. Hence, $\max_{\vec x, \vec y\in\mathcal M_k} \Vert \vec x - \vec y \Vert \rightarrow 0$ as $k\to\infty$. Then, by virtue of \cref{lem:consist} and the observation that \cref{eq:sitbndfirst} and \cref{eq:sitbnd} are equivalent, \ref{alg:sitbb:bounding} satisfies the convergence criterion in \cref{lem:conv}. \Cref{lem:sitfeas} establishes that the point in \ref{alg:sitbb:feaspoint} is primal feasible and suitable with respect to \cref{lem:consist}. It follows that, for fixed $\delta_k$, after a finite number of iterations, either a primal feasible point is found or all boxes are pruned in \ref{alg:sitbb:prune} and the algorithm is terminated in \ref{alg:sitbb:terminate}. Hence, from \cref{lem:duality}, $\delta_k$ is, upon termination, either a $(\varepsilon, \eta)$-optimal solution of \cref{eq:srmax} or, if $\delta_k$ was not set with some primal feasible point, the problem is $\varepsilon$-essential infeasible.

	It is established in \cite[App.~C]{sit} that updating $\delta_k$ with encountered primal feasible points does not invalidate the bounds in $\mathscr R_k$. Hence, restarting the procedure upon updating $\delta_k$ in \ref{alg:sitbb:feaspoint} is not necessary to ensure correct convergence. Finally, observe that the primal objective is bounded above by the global optimum and below by zero. Hence, the initialization of $\delta_0$ in \ref{alg:sitbb:init} is valid. Moreover, the sequence $\{ \delta_k \}_k$ converges to a value between $v\eqref{eq:srmax}$ and $v\eqref{eq:srmax} + \eta$. For $\eta > 0$, this sequence is clearly finite.
\end{IEEEproof}

\section{Numerical Evaluation}
\label{sec:numeval}
\pgfplotsset{
	defaultplot/.style={
		thick,
		ylabel near ticks,
		xlabel near ticks,
		grid=major,
		ymin = 0,
		minor x tick num = 1,
		minor y tick num = 1,
		%yminorgrids = true,
		%mark repeat = 2,
		%xmin = 0,
		%xmax = 24,
		%ymin = 0,
		%ymax = .9,
		legend cell align=left,
		legend style={font=\scriptsize, nodes={inner sep=1pt}, inner xsep=1.5pt, inner ysep=1pt},
		legend image post style={xscale=.7},
		font={\footnotesize},
		legend entries = {RSMA, MU-LP, NOMA},
		cycle list name=default,
		no markers,
	},
	smallplot/.style={
		defaultplot,
		width=.57*\axisdefaultwidth,
	},
	rateregion/.style={
		smallplot,
		xlabel={$C_1 + R_{p,1}$ [bpcu]},
		ylabel={$C_2 + R_{p,2}$ [bpcu]},
		legend pos=south west,
		xmin = 0,
	},
	wsr/.style={
		smallplot,
		xlabel={SNR [dB]},
		ylabel={Sum Rate [bpcu]},
		legend pos=north west,
		xmin = 5,
		xmax = 30,
	},
	ee/.style={
		smallplot,
		xlabel={Power $P$ [dBm]},
		ylabel={Energy Efficiency [bits/J/Hz]},
		ylabel style={font=\scriptsize},
		legend pos=south east,
		xmin = 4,
		xmax = 24,
	},
	wmmse/.style={black, dotted}
}

In this section, we employ the developed algorithm to compare \cgls{rsma} with \cgls{mulp} and \cgls{noma} in terms of achievable rate region, maximum sum rate and \cgls{ee}. Further, the performance of first-order optimal solution methods for these problems is measured against the global solution and some numerical properties of \cref{alg:sitbb} are examined.

\subsection{System Performance}
\label{sec:numeval:system}
Consider a \cgls{bs} with $M=2$ transmit antennas that serves $K=2$ single antenna users on the same spectrum as described in \cref{sec:sysmod}. We employ \cref{alg:sitbb} to solve the beamforming problem \cref{eq:srmaxequiv} for \cgls{rsma} and its special cases \cgls{mulp} and 2-user \cgls{noma}.
The goal of the experiments in this subsection is to evaluate the performance gap between these schemes based on the strong optimality guarantees provided by \cref{alg:sitbb}. In addition, we also obtain beamforming solutions using state-of-the-art first-order optimal algorithms and compare them to the results of \cref{alg:sitbb}. This will give an indication of the usability of those faster algorithms for practical system evaluation.

The channels of user $k$ are chosen randomly using circularly symmetric complex Gaussian distribution with zero mean and variance $\sigma_k^2$. A set of 100 \cgls{iid} feasible channel realizations is generated for each simulation separately. Computation time and memory consumption per problem instance were limited. This leads to results being averaged over less than 100 samples per simulation. Two different channel statistics are considered: one where both users' channels are generated with equal variances and one with roughly \SI{10}{\dB} disparity in the variances.

\subsubsection{Rate Region}
\begin{figure}%
	\tikzsetnextfilename{rr1}%
	\tikzpicturedependsonfile{RateRegion_SNR20NT2bias1.dat}%
	\subfloat[$\sigma_1^2=\sigma_2^2=1$]{%
		\centering
		\begin{tikzpicture}
			\begin{axis} [rateregion]
				\pgfplotstableread[col sep=comma]{RateRegion_SNR20NT2bias1.dat}\tbl

				\addplot table[x=RSMA_R1, y=RSMA_R2] {\tbl};
				\addplot table[x=MULP_R1, y=MULP_R2] {\tbl};
				\addplot table[x=NOMA_R1, y=NOMA_R2] {\tbl};

				\addplot[wmmse] table[x=LO_RSMA_R1, y=LO_RSMA_R2] {\tbl};
				\addplot[wmmse] table[x=LO_MULP_R1, y=LO_MULP_R2] {\tbl};
				\addplot[wmmse] table[x=LO_NOMA_R1, y=LO_NOMA_R2] {\tbl};
			\end{axis}
		\end{tikzpicture}%
		\label{fig:RR1}%
	}
	\hfill%
	\tikzsetnextfilename{rr2}%
	\tikzpicturedependsonfile{RateRegion_SNR20NT2bias03.dat}%
	\subfloat[$\sigma_1^2=1, \sigma_2^2=0.09$]{%
		\centering
		\begin{tikzpicture}
			\begin{axis} [rateregion, yticklabel pos = right]
				\pgfplotstableread[col sep=comma]{RateRegion_SNR20NT2bias03.dat}\tbl

				\addplot table[x=RSMA_R1, y=RSMA_R2] {\tbl};
				\addplot table[x=MULP_R1, y=MULP_R2] {\tbl};
				\addplot table[x=NOMA_R1, y=NOMA_R2] {\tbl};

				\addplot[wmmse] table[x=LO_RSMA_R1, y=LO_RSMA_R2] {\tbl};
				\addplot[wmmse] table[x=LO_MULP_R1, y=LO_MULP_R2] {\tbl};
				\addplot[wmmse] table[x=LO_NOMA_R1, y=LO_NOMA_R2] {\tbl};
			\end{axis}
		\end{tikzpicture}%
		\label{fig:RR2}%
	}%
	\caption{Achievable rate regions for \cgls{rsma}, \cgls{mulp} and \cgls{noma} at an \cgls{snr} of \SI{20}{\dB}. Colored lines are globally optimal results obtained from \cref{alg:sitbb} and dashed lines are the corresponding results from a \cgls{wmmse} algorithm.}
	\label{fig:RR}
\end{figure}
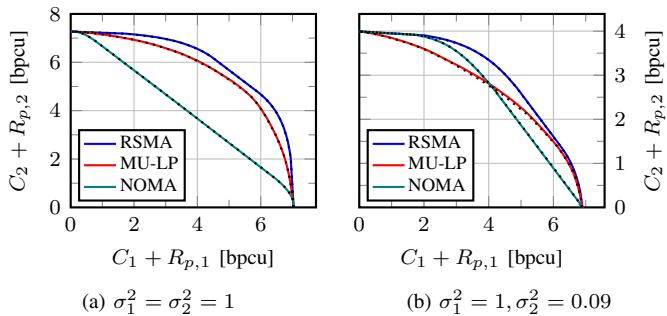

We start with the achievable rate region for \cgls{rsma}, \cgls{mulp} and \cgls{noma}. The boundary points for each strategy are calculated by setting $R_k^{th} = 0$ and $u_1 = 1$ in \cref{eq:WSR QoS prob}. Following \cite{wmmse2008}, we vary the weight $u_2 \in \{10^x | x = -3, -1, -0.95, -0.8, \cdots, 0.95, 1, 3 \}$. The resulting rate region is obtained from the convex hull over the computed boundary points. Results for a \cgls{snr} of \SI{20}{\dB} are displayed in \cref{fig:RR}. \Cref{fig:RR1} was averaged over 63 channel realizations, while \cref{fig:RR2} was obtained from 61 realizations. It can be observed that the achievable rate region of \cgls{rsma} is strictly larger than that of \cgls{mulp} and \cgls{noma}. In case of equal channel statistics, \cgls{mulp} also strictly outperforms \cgls{noma}, while in the case with disparate statistics neither \cgls{mulp} nor \cgls{noma} is superior to the other. However, as \cgls{rsma} includes both strategies as special cases and allows arbitrary combinations of them, its rate region is strictly larger. These observations are in line with previous evaluations \cite{mao2017eurasip} but are scientifically more reliable since they rely on proven globally optimal solutions to \cref{eq:srmaxequiv}.

The first-order optimal solutions are computed using the \cgls{wmmse} algorithm from \cite{mao2017eurasip} for \cgls{rsma} and \cgls{noma}, and \cgls{mulp}. For each parameter combination, a single initialization was used. In particular, the common and private stream precoders are initialized using \cgls{svd} and \cgls{mrt}, respectively, as in \cite{mao2021IoT}.
The results are displayed as black dashed lines. While the \cgls{wmmse} algorithm does not always achieve the optimal solution, as can be seen in the \cgls{mulp} performance in \cref{fig:RR2}, its solution is sufficiently close to the true solution to allow drawing conclusions based on the results. Moreover, the obtained solution is well suited for practical system design.

\subsubsection{Sum Rate Maximization}
\label{sec:numeval:system:sr}

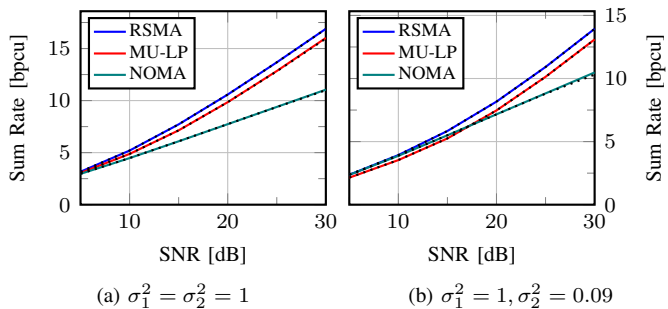
\begin{figure}%
	\tikzsetnextfilename{wsr1}%
	\tikzpicturedependsonfile{WSR_Nt2bias1.dat}%
	\subfloat[$\sigma_1^2=\sigma_2^2=1$]{%
		\centering
		\begin{tikzpicture}
			\begin{axis} [wsr]
				\pgfplotstableread[col sep=comma]{WSR_Nt2bias1.dat}\tbl

				\addplot table[y=RSMA] {\tbl};
				\addplot table[y=MULP] {\tbl};
				\addplot table[y=NOMA] {\tbl};

				\addplot[wmmse] table[y=LO_RSMA] {\tbl};
				\addplot[wmmse] table[y=LO_MULP] {\tbl};
				\addplot[wmmse] table[y=LO_NOMA] {\tbl};
			\end{axis}
		\end{tikzpicture}%
		\label{fig:WSR1}%
	}%
	\hfill%
	\tikzsetnextfilename{wsr2}%
	\tikzpicturedependsonfile{WSR_Nt2bias03.dat}%
	\subfloat[$\sigma_1^2=1, \sigma_2^2=0.09$]{%
		\centering
		\begin{tikzpicture}
			\begin{axis} [wsr, yticklabel pos = right]
				\pgfplotstableread[col sep=comma]{WSR_Nt2bias03.dat}\tbl

				\addplot table[y=RSMA] {\tbl};
				\addplot table[y=MULP] {\tbl};
				\addplot table[y=NOMA] {\tbl};

				\addplot[wmmse] table[y=LO_RSMA] {\tbl};
				\addplot[wmmse] table[y=LO_MULP] {\tbl};
				\addplot[wmmse] table[y=LO_NOMA] {\tbl};
			\end{axis}
		\end{tikzpicture}%
		\label{fig:WSR2}%
	}%
	\caption{Maximum achievable sum rate for \cgls{rsma}, \cgls{mulp} and \cgls{noma} with increasing \cgls{qos} constraints (see text for details). Colored lines are globally optimal results obtained from \cref{alg:sitbb} and dashed lines are the corresponding results from a \cgls{wmmse} algorithm.}
	\label{fig:WSR}
\end{figure}

We maximize the sum rate under \cgls{qos} constraints, i.e., we solve \cref{eq:WSR QoS prob} with $u_k = 1$, $k = 1, 2$. A \cgls{snr} range from \SI{5}{\dB} to \SI{30}{\dB} with \SI{5}{\dB} increments is considered. The corresponding \cgls{qos} constraints $R_k^{th}$ are chosen as 0.1, 0.2, 0.4, 0.6, 0.8, and 1 (all in \si{\bpcu}), respectively. The results are displayed in \cref{fig:WSR}. Both plots were obtained by averaging over 90 \cgls{iid} channel realizations. 
%The same observations as before hold true: \cgls{rsma} outperforms \cgls{mulp} and \cgls{noma} in both cases, while there is not a clearly superior strategy when comparing \cgls{mulp} and \cgls{noma}.
	When there is no channel strength disparity, as in \cref{fig:WSR1}, \cgls{noma} achieves the worst sum rate performance among the considered schemes. This is not unexpected as \cgls{noma} exploits channel strength disparities among users.
	 When there is a \SI{10}{\dB} channel strength difference between the two users, \cgls{noma} slightly outperforms \cgls{mulp} in the low \cgls{snr} regime. This is likely caused by the minimum rate constraints, which move the operating point towards a solution that plays to the strengths of \cgls{noma}, as can be seen from the rate region in \cref{fig:RR2}.
	Starting from approximately \SI{17}{\dB}, \cgls{mulp} clearly outperforms \cgls{noma}. A likely reason for this is that the additional decoding constraint in \cgls{noma} results in a reduced spatial multiplexing gain at high \cglspl{snr} \cite{bruno2021MISONOMA}.
Thanks to the ability of partially decoding interference and partially treat interference as noise, \cgls{rsma} combines the strengths of \cgls{noma} and \cgls{mulp} and, thus, outperforms them in both cases.
Again, the \cgls{wmmse} algorithm performs quite well compared to the global solution and appears to be a good practical choice. % in most cases.

\subsubsection{Energy Efficiency}
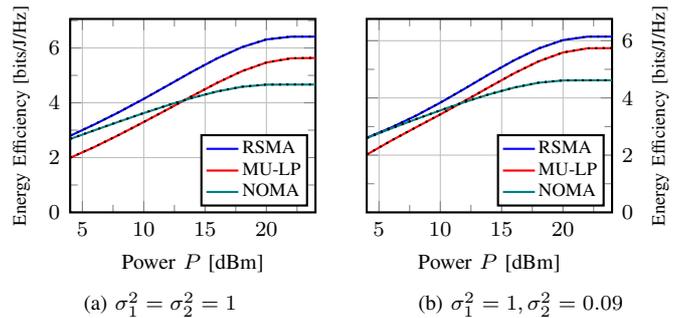
\begin{figure}%
	\tikzsetnextfilename{ee1}%
	\tikzpicturedependsonfile{EE_Nt2bias1QoS1Pnoise1mW.dat}%
	\subfloat[$\sigma_1^2=\sigma_2^2=1$]{%
		\centering
		\begin{tikzpicture}
			\begin{axis} [ee]
				\pgfplotstableread[col sep=comma]{EE_Nt2bias1QoS1Pnoise1mW.dat}\tbl

				\addplot table[y=RSMA] {\tbl};
				\addplot table[y=MULP] {\tbl};
				\addplot table[y=NOMA] {\tbl};

				\addplot[wmmse] table[y=LO_RSMA] {\tbl};
				\addplot[wmmse] table[y=LO_MULP] {\tbl};
				\addplot[wmmse] table[y=LO_NOMA] {\tbl};
			\end{axis}
		\end{tikzpicture}%

		\label{fig:EE1}%
	}%
	\hfill%
	\tikzsetnextfilename{ee2}%
	\tikzpicturedependsonfile{EE_Nt2bias03QoS1Pnoise1mW.dt}%
	\subfloat[$\sigma_1^2=1, \sigma_2^2=0.09$]{%
		\centering
		\begin{tikzpicture}
			\begin{axis} [ee, yticklabel pos = right]
				\pgfplotstableread[col sep=comma]{EE_Nt2bias03QoS1Pnoise1mW.dat}\tbl

				\addplot table[y=RSMA] {\tbl};
				\addplot table[y=MULP] {\tbl};
				\addplot table[y=NOMA] {\tbl};

				\addplot[wmmse] table[y=LO_RSMA] {\tbl};
				\addplot[wmmse] table[y=LO_MULP] {\tbl};
				\addplot[wmmse] table[y=LO_NOMA] {\tbl};
			\end{axis}
		\end{tikzpicture}%
		\label{fig:EE2}%
	}%
	\caption{Energy efficiency of \cgls{rsma}, \cgls{mulp} and \cgls{noma} with $R_k^{th} = \SI{1}{\bpcu}$. Colored lines are globally optimal results obtained from \cref{alg:sitbb} and dashed lines are the corresponding results from an \cgls{sca} algorithm.}
	\label{fig:EE}
\end{figure}

The third problem type supported by \cref{alg:sitbb} is \cgls{ee} maximization. We solve problem \cref{eq:EE prob} for maximum transmit powers ranging from \SI{4}{\dBm} to \SI{30}{\dBm} in steps of \SI{2}{\dBm}, with power amplifier inefficiency $\mu = 0.35$, noise variance $0.0001$, and static circuit power consumption $P_c = M P_\mathrm{dyn} + P_\mathrm{sta}$, where $M$ is the number of transmit antennas,  $P_\mathrm{dyn} = \SI{27}{\dBm}$, and $P_\mathrm{sta} = \SI{1}{\milli\W}$.
Potentially suboptimal solutions are computed with the \cgls{sca} approach \cite{IETBook} as in \cite{mao2018EE} for  \cgls{rsma}, \cgls{noma}, and \cgls{mulp}. Again a single initialization per parameter combination is used, following the same methodology as in \cite{mao2018EE}.
Results are shown in \cref{fig:EE1,fig:EE2} and were obtained by averaging over 29 and 32 \cgls{iid} channel realizations, respectively. The results follow the usual shape of \cgls{ee} maximization, where the \cgls{ee} first increases and then saturates at some point. Interestingly, while \cgls{rsma} clearly outperforms the other two schemes, \cgls{mulp} has always higher \cgls{ee} than \cgls{noma} when the transmit power budget is large enough, while for constrained transmit powers, \cgls{noma} has slightly higher efficiency than \cgls{mulp}.
This is because the \cgls{ee} follows the sum rate performance until it saturates.
As before, the first-order optimal results, this time obtained with \cgls{sca}, are very close to the globally optimal solution and we can conclude that in most cases such an algorithm will be sufficient for performance analysis.

\subsection{Numerical Performance}
We have evaluated \cref{alg:sitbb} and two state-of-the-art first-order optimal methods in a real world setting. The key observation is that the \cgls{wmmse} and \cgls{sca} methods without proven convergence to the global solution perform very well and, on average, are virtually equal to the globally optimal solution. In this subsection, we first take a closer look at the numerical accuracy of the first-order optimal methods and then study the numerical stability and complexity of \cref{alg:sitbb}.

\subsubsection{Numerical Accuracy}
\begin{figure}%
	\tikzsetnextfilename{cdf}%
	\tikzpicturedependsonfile{cdf.dat}%
	\centering
	\begin{tikzpicture}
		\begin{axis} [
				width=\axisdefaultwidth,
				height=.7*\axisdefaultheight,
				defaultplot,
				legend pos = north west,
				ymin = -0.05,
				ymax = 1.05,
				xmin = -0.1,
				xmax = 0.115,
				extra x ticks = {0.02},
				extra x tick labels = {$\eta$},
				tick label style={/pgf/number format/fixed},
				ylabel = {Cumulative Probability},
				xlabel = {$f(\bar{\vec x}) - f(\tilde{\vec x})$},
			]
			\pgfplotstableread[col sep=comma]{cdf.dat}\tbl

			\addplot table[y=RSMA] {\tbl};
			\addplot+[dashed] table[y=MULP] {\tbl};
			\addplot+[densely dotted] table[y=NOMA] {\tbl};
		\end{axis}
	\end{tikzpicture}
	\caption{Empirical \cgls{cdf} of the difference between the optimal values returned by \cref{alg:sitbb} and the first-order optimal baseline algorithm, where $f(\bar{\vec x})$ is the optimal value returned by \cref{alg:sitbb} and $f(\tilde{\vec x})$ is the corresponding objective value for the solution returned by the first-order optimal solver.}
	\label{fig:cdf}
\end{figure}
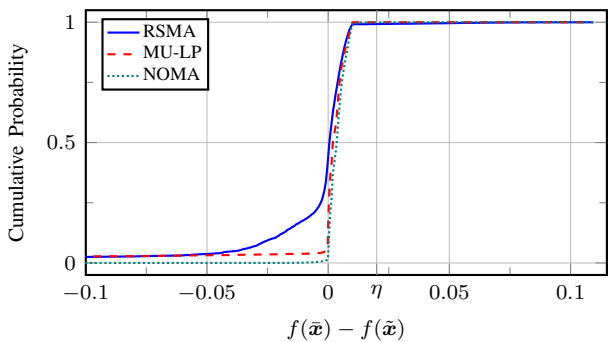

The results in \cref{sec:numeval:system} were obtained from \cref{alg:sitbb} with tolerances $\eta = 0.02$ and $\varepsilon = 10^{-7}$. The numerical tolerances of the first-order optimal methods were chosen small enough not to be relevant. \Cref{fig:cdf} shows the empirical \cgls{cdf} of the difference between the globally optimal solution and the first-order optimal solution computed for the analyses in \cref{sec:numeval:system} with both metrics, \cgls{wsr} and \cgls{ee}. Accordingly, a total of \num{11520} computed data points for \cgls{rsma}, \num{12600} points for \cgls{mulp} and \num{24880} for \cgls{noma} form the basis of \cref{fig:cdf}. A negative value indicates that the solution of \cref{alg:sitbb} achieves a larger objective value than that computed by the \cgls{wmmse} or \cgls{sca} approach. Instead, a positive value indicates that the first-order optimal solution is better than the one obtained by \cref{alg:sitbb}. Recalling the definition of $\eta$-optimality in \cref{eq:etaopt}, it is apparent that this is not an unexpected outcome.

There exists a small amount of solutions returned by \cref{alg:sitbb} that is not within an $\eta$-region around the global optimal solution. This is indicated by a deviation of more than $\eta$ from the first-order optimal solution. In particular, this affects 98 of the \cgls{rsma} solutions (\SI{0.85}{\percent}), 40 of the \cgls{noma} solutions (\SI{0.16}{\percent}), and none of the \cgls{mulp} solutions. The reason for this is the tightening of nonconvex constraints necessary for the \cgls{sit} approach. Reducing the size of $\varepsilon$ will resolve this numerical issue but also leads to slower convergence. The fact that the \cgls{mulp} solutions are unaffected indicates that the likely reason is in the tightening of \cref{eq:srmax:3,eq:srmax:8,eq:srmax:4}.

\begin{figure}%
	\tikzsetnextfilename{cdf3}%
	\tikzpicturedependsonfile{cdf_rsma3.dat}%
	\tikzpicturedependsonfile{cdf_mulp3.dat}%
	\centering
	\begin{tikzpicture}
		\begin{axis} [
				smallplot,
				legend pos = north west,
				ymin = -0.05,
				ymax = 1.05,
				xmax = 0.25,
				xminorgrids = true,
				tick label style={/pgf/number format/fixed},
				ylabel = {Cumulative Probability},
				xlabel = {$f(\bar{\vec x}) - f(\tilde{\vec x})$},
			]

			\addplot table[y=RSMA,col sep=comma] {cdf_rsma3.dat};
			\addplot+[dashed] table[y=MULP,col sep=comma] {cdf_mulp3.dat};
		\end{axis}
	\end{tikzpicture}
	\tikzsetnextfilename{cdf32}%
	\begin{tikzpicture}
		\begin{axis} [
				smallplot,
				yticklabel pos = right,
				legend pos = north west,
				tick label style={/pgf/number format/fixed},
				scaled x ticks = false,
				ymin = -0.05,
				ymax = 1.05,
				xmin = -0.05,
				xmax = 0.01,
				extra x ticks = {0.01},
				extra x tick labels = {$\eta$},
				ylabel = {Cumulative Probability},
				xlabel = {$f(\bar{\vec x}) - f(\tilde{\vec x})$},
			]

			\addplot table[y=RSMA,col sep=comma] {cdf_rsma3.dat};
			\addplot+[dashed] table[y=MULP,col sep=comma] {cdf_mulp3.dat};
		\end{axis}
	\end{tikzpicture}%
	\caption{Empirical \cgls{cdf} of the difference between globally optimal and first-order optimal solution for 3-user sum rate maximization. Same display style as in \cref{fig:cdf}. Complete value range on the left, relevant value range around zero on the right.}
	\label{fig:cdf3}
\end{figure}
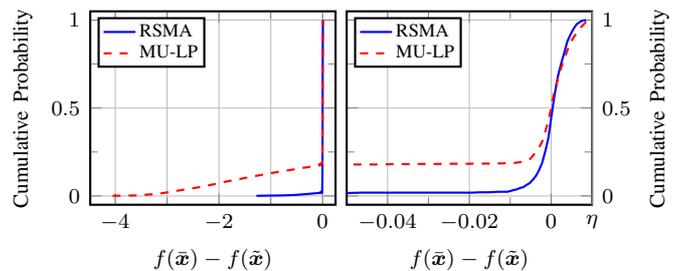

A relevant question is whether this virtually optimal behavior of first-order optimal methods continues for more than two users. To this effect, we have considered sum rate maximization for a scenario with three users and four antennas. Parameters are selected as in \cref{sec:numeval:system:sr}. Channels are generated with unit variance as before. The \cgls{cdf} of the absolute error between the result from \cref{alg:sitbb} and the \cgls{wmmse} algorithm from \cite{mao2017eurasip} is shown in \cref{fig:cdf3}. For each algorithm, \num{1005} data points have been considered, each with random channel initialization and one of the (\cgls{snr}, $R_{th}$) combinations from \cref{sec:numeval:system:sr}. The performance of the \cgls{wmmse} algorithm is similar to the 2-user case for \cgls{rsma}. Interestingly, this behavior is not shown for \cgls{mulp} precoding, where a relevant number of \cgls{wmmse} results is considerably worse than the globally optimal solution.

\subsubsection{Numerical Stability}
Next, we evaluate the numerical stability of \cref{alg:sitbb} in comparison to conventional \cgls{bb} based methods. We focus on multiple unicast beamforming, i.e., where $\vec p_c = \vec 0$, as this special case of the more general problem is already quite difficult to solve with \cgls{bb}.
As baseline comparison, we implemented two methods. The first is a straightforward \cgls{bb} solution of \cref{eq:srmax} with $\vec p_c = 0$ as published in \cite{Bjornson2013,Tervo2015}. We denote this algorithm as ``BB'' in the results. The bounding problem in this algorithm is often difficult to solve for state-of-the-art convex solvers (e.g., Mosek \cite{mosek}) since the feasible region can become extremely small. Following \cite[\S 2.2.2]{Bjornson2013}, the feasible set can be relaxed such that the bounding problem always has good numerical properties. Interestingly, the resulting problem is similar to the \cgls{sit} bounding problem in \cref{sec:sit:bounding}. The downside of this approach is that feasible point acquisition for the \cgls{bb} procedure becomes much harder. We denote this algorithm as ``BB2''.

The numerical evaluation is based on 100 random \cgls{iid} channel realizations. We solved \cref{eq:srmaxequiv} for $u_k = 1$, $\vec p_c = \vec 0$, $\mu = 0$, $P_c = 0$, $R_{k}^{th} = 0$, $\frac{P}{\mathrm{dB}} = -10, -5, \dots, 20$, and $K = M \in \{ 2, 3, 4 \}$. This results in 700 problem instances per $K$.

For $K = 2$, BB2 stalled in 364 problem instances, while the other algorithms solved all problems. For $K = 3$, BB2 stalled in 146 instances and BB failed 13 times due to numerical problems of the convex solver. Finally, for $K = 4$, BB did not solve a single problem instance due to numerical issues and BB2 stalled in 27 instances. Moreover, \cref{alg:sitbb} and BB2 did not solve the problem within 60 minutes in 4 and 60 instances, respectively. Average computation times on a single core of an Intel Cascade Lake Platinum 9242 CPU are reported in \cref{tab}. It can be observed that the proposed \cref{alg:sitbb} is more efficient than the two baseline algorithms especially when more users are in the system.
Moreover, the joint beamforming problem, i.e., with $\vec p_c \neq \vec 0$, was solved by \cref{alg:sitbb} for $K = 2$ with mean and median run times of \SI{942}{\second} and \SI{2786}{\second}. However, 23 instances were not solved within 12 hours and in the simulations presented in \cref{sec:numeval:system}, we observed some parameter combinations with very slow convergence speed, especially in \cgls{ee} maximization problems.

%As a final note, observe from the discussion in \cref{sec:alg} that the complexity scales with $O(\exp(2 K))$ in the number of users and polynomially in the number of antennas $M$. Hence, no noticeable changes in the reported run times are to be expected by varying $M$.

\begin{table}[tb]
	\caption{Mean and median run times to obtain the optimal solution for MU-LP precoding. Instances where not all algorithms converged are ignored.}
	\label{tab}
	\centering
	\footnotesize
	\begin{tabular}{lccc}
		\toprule
		 & $K = 2$ & $K = 3$ & $K = 4$\\
		 \midrule
		 Alg.~\ref{alg:sitbb} & \SI{0.175}{\second} / \SI{0.099}{\second} & \SI{4.579}{\second} / \SI{1.959}{\second} & \SI{334.8}{\second} / \SI{126.3}{\second} \\
		 BB  & \SI{0.173}{\second} / \SI{0.091}{\second} & \SI{7.605}{\second} / \SI{2.606}{\second} & --- \\
		 BB2 & \SI{42.41}{\second} / \SI{2.380}{\second} & \SI{158.5}{\second} / \SI{12.42}{\second} & \SI{704.1}{\second} / \SI{265.8}{\second}\\
		\bottomrule
	\end{tabular}
\end{table}

\subsubsection{Numerical Complexity}
Finally, we discuss the numerical complexity of \cref{alg:sitbb}. As mentioned in \cref{sec:sitfund}, the \cgls{sit} framework has exponential complexity in the number of variables. \Cref{alg:sitbb} applies this \cgls{sit} approach to the variables $\vec \gamma_p, s, \vec\alpha$, while all other variables of \cref{eq:srmax} are treated in the subproblems for bounding, feasible point determination, etc. The number of ``\cgls{sit} variables'' scales as $2 K$, while the dimension of the subproblems scales as $\mathcal O(K M)$. Since all subproblems are convex optimization problems that can be solved in polynomial time, the complexity of solving each subproblem is polynomial in the number of antennas $M$ and the number of users $K$. Thus, \cref{alg:sitbb} scales polynomially in $M$ and exponentially in $K$. This theoretical behavior is verified numerically in \cite{sit} for a similar algorithm.

This establishes that \cref{alg:sitbb} is not applicable to scenarios with a large number of users $K$, but scales well with the number of antennas $M$. However, it remains to evaluate how many users are actually manageable with \cref{alg:sitbb}. For the MU-LP case, a partial answer is provided in \cref{tab} in so far that at least four users are feasible with reasonable average run time of a few minutes. Unfortunately, this behavior does not carry over to the \cgls{rsma} case.

The run time for the sum rate maximization in \cref{sec:numeval:system:sr} was, on average, \SI{3.28}{\hour} with standard deviation \SI{11.61}{\hour} and median \SI{0.37}{\hour} for the 2-user case. For three users, the mean run time was \SI{54.72}{\hour} with standard deviation \SI{26.42}{\hour} and median \SI{54.12}{\hour}. In both cases, the computation was terminated prematurely for some channel instances after a maximum run time of \SI{173.76}{\hour} and \SI{102.98}{\hour}, respectively.
The implications of this early termination are twofold. First, the reported values would be higher if computation for all channel realizations would have been completed. Second, the spread between the two and three users cases would be even larger, as the three user results were terminated earlier than the two users case.
For the rate region results, similar numbers can be reported, while the \cgls{ee} maximization ran significantly longer.

Clearly, computing precoding solutions for \cgls{rsma} has a considerably higher effective numerical complexity than the \cgls{mulp} scenario.
We can only speculate on the reasons for this. One is, for sure, that \cgls{rsma} has twice as many nonconvex variables, i.e., $2K$ instead of $K$ for the \cgls{mulp} scenario. However, this does not completely account for the larger run times, as the reported times for four variables in \cref{tab} are still much lower than for the 2-user \cgls{rsma} scenario. This leaves only the conclusion that the multicast precoding part of the \cgls{rsma} solution is much harder to compute than the unicast part. Potential approaches to improve the performance of this part are replacing/improving the argument cut strategy and additional reduction steps for the argument cut variables. Similarly, our hypothesis for improving the performance of \cgls{ee} maximization is a different reduction approach. These potential performance improvements are left open for future work.

\section{Conclusions}
\label{sec:conclusions}
We have developed a globally optimal beamforming algorithm for \cgls{wsr} and \cgls{ee} maximization in \cgls{miso} downlink systems with \cgls{rsma}. The algorithm exhibits finite convergence and is the first method to solve this optimization problem. It is also the first beamforming algorithm based on the \cgls{sit}-\cgls{bb} approach.
Two user \cgls{noma} and \cgls{mulp} beamforming are incorporated as special cases. We have shown numerically that the proposed algorithm outperforms state-of-the-art globally optimal beamforming algorithms for the \cgls{mulp} problem, both in terms of numerical stability and practical convergence speed. Extensive numerical experiments establish that contemporary suboptimal solution methods for \cgls{rsma} beamforming often obtain a solution very close to the global optimum. In particular, there is virtually no difference between the suboptimal solution and the true optimum when evaluating the average performance over a large number of channel realizations. Hence, this paper establishes that \cgls{wmmse} and \cgls{sca}-based methods are  suitable choices for such performance comparisons, at least in the 2-user case. This effectively strengthens the results of many earlier studies in this area, as it retrospectively validates the numerical approach taken to compare the performance of \cgls{rsma} against \cgls{noma} and \cgls{mulp}.

\appendices
\section{Proof of \cref{prop:srmax}} \label{proof:prop:srmax}
Let $(\vec x^\star, \vec \gamma_c^\star)$ be a solution of \cref{eq:srmaxequiv} and set $s^\star = \min_k \gamma_{c,k}^\star$. Constraints~\cref{eq:srmaxequiv:4,eq:srmaxequiv:6} are part of both problems. Constraint~\cref{eq:srmaxequiv:3} is equivalent to
	\begin{equation} \label{eq:srmaxproof:1}
		\sum_{k\in \mathcal{K}}C_{k}\leq \log(1+ \min_{k\in\mathcal K} \gamma_{c,k}).
	\end{equation}
	Since $(\vec x^\star, \vec \gamma_c^\star)$ satisfies \cref{eq:srmaxproof:1}, $(\vec x^\star, s^\star)$ satisfies \cref{eq:srmax:9}. Finally, \cref{eq:srmax:2} is a relaxed version of \cref{eq:srmaxequiv:2} and \cref{eq:srmax:3} is equivalent to the definition of $s^\star$.

	For the converse, let $(\vec x^\star, s^\star)$ be a solution of \cref{eq:srmax} and set $\gamma_{c,k}^\star = \frac{|{\vec{h}}_{k}^{H}\vec{p}_{c}^\star|^2}{\sum_{j\in\mathcal{K}}|\vec{h}_{k}^{H}\vec{p}_{j}^\star|^2+1}$ for all $k\in\mathcal K$.
	Since the objective is increasing in $\vec\gamma_p$, constraint \cref{eq:srmaxinter:1} is always active in the optimal solution if $u_k > 0$. Otherwise, i.e., for $u_k = 0$, relaxing \cref{eq:srmaxequiv:2} does not relax \cref{eq:srmaxequiv:4}. Hence, \cref{eq:srmaxinter:1} is equivalent to the $\gamma_{p,k}$ part of \cref{eq:srmaxequiv:2} (and the $\gamma_{c,k}$ part is satisfied by definition).
	Constraint~\cref{eq:srmaxequiv:3} is equivalent to \cref{eq:srmaxproof:1}. With an auxiliary variable $s = \min_k \gamma_{c,k}$, the \cgls{rhs} of \cref{eq:srmaxproof:1} can be replaced by $\log(1+s)$. Due to monotonicity, the relaxed version $s \le \min_k \gamma_{c,k}$ is active in the optimal solution. Observing that \cref{eq:srmaxinter:2} is the smooth variant of this constraint completes this part of the proof.

	For the last part of the proposition, it suffices to show that every solution of \cref{eq:srmax} solves \cref{eq:srmaxinter}.
Observe that \cref{eq:srmaxinter:1} is equivalent to
\begin{equation} \label{eq:srmaxproof:2}
	\sqrt{\gamma_{p,k}} \left( \sum\nolimits_{j\in\mathcal{K}\setminus k}|\vec{h}_{k}^{H}\vec{p}_{j}|^2+1 \right)^{1/2} \le |{\vec{h}}_{k}^{H}\vec{p}_{k}|
\end{equation}
and that the solution is invariant to rotations of $\vec p_k$, $k\in\mathcal K$, i.e., if $\vec p_k^\star$ solves \cref{eq:srmaxinter}, then $\vec p_k^\star e^{j\phi}$ also solves \cref{eq:srmaxinter} for all real-valued $\phi$ \cite{Bengtsson1999}. Hence, constraint \cref{eq:srmax:4} can be added to \cref{eq:srmaxinter} without reducing the optimal value. Then, \cref{eq:srmaxproof:2} is equivalent to \cref{eq:srmax:1}.

Similarly, \cref{eq:srmaxinter:2} is equivalent to
\begin{equation} \label{eq:srmaxproof:3}
	\sqrt s \left( \sum\nolimits_{j\in\mathcal{K}}|\vec{h}_{1}^{H}\vec{p}_{j}|^2+1 \right)^{1/2} \le |{\vec{h}}_{k}^{H}\vec{p}_{c}|
\end{equation}
for all $k\in\mathcal K$ and the solution is invariant to rotations in $\vec p_c$. However, except for degenerate cases, only one \cgls{rhs} of \cref{eq:srmaxproof:3} can be made real-valued. \cGls{wlog}, this is done for $k = 1$ by adding \cref{eq:srmax:5} to \cref{eq:srmaxinter}. For the remaining $K-1$ constraints, introduce auxiliary variables $d_k = |\vec h_k^H \vec p_c|$ and observe that relaxing $0 \le d_k \le |\vec h_k^H \vec p_c|$ does not decrease the optimal value of \cref{eq:srmaxinter}. However, it also does not increase the optimal value since it is increasing in $s$ and, hence, also increasing in $d_k$. Finally, introducing the constraint $e_k = \vec h_k^H \vec p_c$ results in \cref{eq:srmax}.

\balance
\bibliography{IEEEabrv,IEEEtrancfg,bibliography.bib}

\end{document}